\documentclass[amsmath,12pt,amssymb,preprint,prd,aps,nofootinbib]{revtex4}
\usepackage{amsfonts} %\usepackage{citesort}
\usepackage{graphicx} % Include figure files
\usepackage{epsfig}
\usepackage{multirow}
\usepackage{bm}% bold math

\begin{document}
\title{Charm mesons in magnetized nuclear matter\\
-- effects of (inverse) magnetic catalysis}
\author{Sourodeep De}
\email{sourodeepde2015@gmail.com}
\author{Pallabi Parui}
\email{pallabiparui123@gmail.com}
\author{Amruta Mishra}
\email{amruta@physics.iitd.ac.in}
\affiliation{Department of Physics, Indian Institute of Technology, Delhi, Hauz Khas, New Delhi - 110 016, India}

\begin{abstract}
We investigate the in-medium masses of the pseudoscalar $(D,{\bar D},D_s^{\pm})$, and vector $(D^*,\bar{D}^*, D_s^{*\pm})$, open charm mesons in isospin asymmetric magnetized nuclear matter, accounting for the effects of Dirac sea. The masses are used to study the in-medium partial decay widths of $D^* \rightarrow D \pi$ ($\bar{D}^*\rightarrow \bar{D}\pi$) and $\Psi(3770) \rightarrow D \bar{D}$, using the $^3P_0$ model. The in-medium masses of the open charm mesons are calculated  from their interactions with the nucleons and scalar mesons within the generalized chiral effective model, in terms of the scalar and number densities of nucleons and the scalar fields fluctuations. The free energy of the magnetized vacuum with Landau energy levels and anomalous magnetic moments (AMMs) of the charged fermions in the single fermion energies are taken into account in the Dirac sea contribution. The effects of Landau energy levels of protons and AMMs of the nucleons are also considered in the magnetized nuclear matter. The light quark condensates are modified considerably with magnetic field, leading to (inverse) magnetic catalysis due to the magnetized Dirac sea effects. The magnetic field causes modifications to occur due to the mixing of the pseudoscalar and longitudinal component of the vector mesons, along with the lowest Landau level contribution to the ground state energy of the charged mesons as point particle correction. For the charmonium state $\Psi(3770)$, the effects of the magnetized Dirac sea are incorporated to the mass modifications through the medium modified scalar dilaton field $\chi$ within the chiral model. The in-medium masses and decay widths of the open charm and charmonium mesons thus obtained should have important observable consequences in the production of the open charm mesons and 
charmonia in peripheral ultra-relativistic heavy ion collision experiments, where huge magnetic fields are expected to be created.

\end{abstract}
\maketitle

\section{Introduction}
\label{sec1}

The study of the hadron properties under extreme conditions of matter is of great relevance in the high energy heavy ion collision experiments and in some astrophysical objects like the magnetars, neutron stars etc.
The heavy flavor mesons, created in the initial stages of these collisions may have impact from the strong magnetic field. As huge magnetic fields are estimated to be generated in the peripheral ultra relativistic heavy ion collision experiments at LHC in CERN and at RHIC in BNL \cite{skokov, fukushima, deng, tuchin, kharzeev}, in the early phase of the collisions. However, there is still an open question on the time evolution of the produced magnetic field due to the complexity in the estimation of the electrical conductivity within the medium. The effects of magnetic field on the in-medium properties of hadrons thus initiate a new area of research with great relevance in the physics of the heavy ion collisions \cite{Hosaka}.

The experimental observables of the relativistic heavy ion collision experiments are affected by the medium modifications of the hadrons. In the presence of  magnetic field, there can be (decrement) enhancement of the light quark condensates with increasing magnetic field, an effect called (inverse) magnetic catalysis \cite{kharzeevmc,kharmc1,chernodub,Preis,menezes,Shovkovy}. In the literature, there have been studies on the effect of  (inverse) magnetic catalysis on quark matter in the context of NJL model \cite{Preis, menezes, ammc, lemmer, guinjl}. The effects of Dirac sea at finite magnetic field have been studied on the vacuum to nuclear matter phase transition within the Walecka model and an extended linear sigma model \cite{haber}. The (inverse) magnetic catalysis effect has been studied in \cite{arghya}, in the context of nuclear matter, accounting for the finite anomalous magnetic moments of the nucleons. The effects of the anomalous magnetic moments of the nucleons have been taken into account through the weak-field expansion of the fermionic propagators in the calculation of the nucleonic one-loop self energy functions. 

Several studies exist on the in-medium properties of the heavy flavor mesons under the various conditions of matter, for e.g., density, temperature and magnetic fields. Various methodologies are employed to study them as illustrated below. One of them is the chiral effective model \cite{amdmeson,amarindamprc,amarvdmesonTprc,amarvepja,DP_AM_Ds,DP_AM_bbar,DP_AM_Bs,AM_DP_upsilon,dmeson_mag,bmeson_mag,charmonium_mag}. Other approaches include the QCD sum rules (QCDSR) \cite{qcdsrc, qcdsrb, gubler, cho91, cho113}, the coupled channel approach \cite{molina}, the Quark meson coupling (QMC) model \cite{krein, thomas}, etc. The $D (\bar D)$ and $D^* (\bar{D}^*)$ mesons are the pseudoscalar and the vector open charm mesons, respectively, which comprise of a heavy charm (c) quark (antiquark) and a light (u,d) antiquark (quark). In the presence of a strange quark flavor (s), the mesons are $D_s$ mesons. The in-medium properties of the open heavy flavor (charm and bottom) mesons and heavy quarkonium states have been studied within the chiral effective model framework \cite{amdmeson,amarindamprc,amarvdmesonTprc,amarvepja,DP_AM_Ds,DP_AM_bbar,DP_AM_Bs,AM_DP_upsilon}, in the absence of magnetic field. In the presence of a magnetic field, the masses of the open and hidden heavy flavour mesons have been studied within the chiral model \cite{dmeson_mag,bmeson_mag,charmonium_mag} and their effects on the in-medium partial decay widths of the charmonium states to $D\bar D$ \cite{charmdecay_mag,charmdw_mag,open_charm_mag_AM_SPM} and bottomonium states to $B\bar B$ \cite{upslndw_mag}. At finite magnetic field, there is mixing phenomena due to the interaction between the pseudoscalar and the longitudinal component of the vector mesons called the PV mixing, which leads to a decrease (increase) in the masses of the pseudoscalar (longitudinal component of the vector) mesons with increasing magnetic field. The PV mixing effects have been studied for the open heavy flavor mesons and heavy quarkonia in \cite{open_charm_mag_AM_SPM, upslndw_mag, Machado, gubler, cho91, cho113, charmdw_mag, suzuki, Iwasaki, qcdsrc, qcdsrb, Alford}. For the charged mesons, there are additional contributions from the Landau energy levels in presence of an external magnetic field. The in-medium partial decay widths of the heavy quarkonia ($\bar{q}q;q=c,b$) going to open heavy flavor mesons have been studied using a field theoretic model for composite hadrons with quark (and antiquark) constituents \cite{charmdw_mag, upslndw_mag, open_charm_mag_AM_SPM}, as well as within the $^3P_0$ model \cite{charmdecay_mag}, in presence of an external magnetic field. The in-medium hadronic decays of the vector open charm mesons, $D^*\rightarrow D\pi$ ($\bar{D}^*\rightarrow \bar{D}\pi$) have been studied incorporating the PV mixing (and the lowest Landau level contribution for the charged mesons) effects for $D-D^{*}$ ($\bar{D}-\bar{D}^{*}$) mesons  at finite magnetic field \cite{open_charm_mag_AM_SPM}. The charm flavored mesons, which are created in the early phase of the heavy ion (peripheral) collisions when the magnetic field can still be large, are important tools in probing the effects of  magnetic field.

In the present work, we investigate the masses of the pseudoscalar and vector open charm ($D$ ($\bar D$), $D^*$ ($\bar{D}^*$)), as well as the charmed, strange ($D_s$, $D_s^{*}$) mesons, in the magnetic nuclear matter, incorporating the magnetized Dirac sea contributions within a chiral effective model. The in-medium masses of the charmonium state, $\Psi(3770)$ are also calculated within the chiral effective model including the Dirac sea effects, to find its in-medium partial decay widths to open charm mesons $D\bar{D}$. The partial decay widths of the vector meson states $D^*(\bar{D}^*)$ to the pseudoscalar meson states $D(\bar{D})$ and a $\pi$ meson, are also studied considering the effects of the magnetized Dirac sea on their masses, using the $^3P_0$ model \cite{3p0a,3p1, 3p0d,3p0e,3p0f,3p0g,3p0h,Friman}. In the $^3P_0$ model, a light quark-antiquark pair is created in the $^3P_0$ state (with quantum numbers similar to  vacuum, $J^{PC} = 0^{++}$), and this light quark (antiquark) combines with the heavy charm antiquark (quark) of the decaying $D^*$ meson at rest, resulting in the production of the open charm $D$ meson and a light flavored $\pi$ meson \cite{Friman,amarvepja}. The matrix element for the decay of the $D^*$ meson state depends on the center-of-mass momentum $|\textbf{p}|$ of the outgoing particles and on the internal structure of the meson states. 

 We organize the paper as follows: In section \ref{sec2}, we discuss in detail the chiral effective model to calculate the in-medium masses of the open charm  and charmonium, $\Psi(3770)$ mesons in the magnetized nuclear matter accounting for the Dirac sea effects. The effects of the point particle correction to the ground state mass of the charged mesons are discussed. The pseudoscalar-vector mesons (PV) mixing in presence of an external magnetic field are obtained using an effective Lagrangian approach and the spin-magnetic field interaction Hamiltonian (only for $D_s$ mesons). In section \ref{sec3}, using the $^3P_0$ model the hadronic decays of the  vector mesons $D^*$ and $\Psi(3770)$ are described. In section \ref{sec4}, the results of the current investigation are discussed. In section \ref{sec5}, we summarize the findings of this work.

\section{In-medium masses of charmonium and open charm mesons}
\label{sec2}
In-medium masses of the pseudoscalar open charm ($D$) mesons and the excited state of charmonium $\Psi(3770)$, in the magnetized nuclear matter are computed within the effective chiral model. The model is based on the non-linear realization of chiral SU(3)$_L\ \times$ SU(3)$_R$ symmetry \cite{coleman, weinberg, bardeen} and QCD scale symmetry breaking effect \cite{Papa, AM2}. The latter is realized through a logarithmic potential in terms of the scalar dilaton field, $\chi$ \cite{erikk}. General structure of the  chiral model Lagrangian density is 
\begin{equation}
{\cal L} = {\cal L}_{kin} + \sum_ W {\cal L}_{BW}
          +  {\cal L}_{vec} + {\cal L}_0
+ {\cal L}_{scale-break}+ {\cal L}_{SB}+{\cal L}_{mag},
\label{genlag} \end{equation}
where, $ {\cal L}_{kin} $ corresponds to the kinetic energy terms
of the baryons and the mesons,
${\cal L}_{BW}$ describes the baryon-meson interactions for $W=$ scalar, pseudoscalar, vector, axial vector mesons,
$ {\cal L}_{vec} $ gives the dynamical mass generation of the vector mesons due to the couplings to the scalar fields
and contains additional quartic vector mesons self-interactions, ${\cal L}_0 $ expresses the meson-meson interaction terms,
${\cal L}_{scale-break}$ is the scale symmetry breaking logarithmic potential in terms of a scalar dilaton field, $\chi$ and $ {\cal L}_{SB} $ illustrates the explicit chiral symmetry
breaking term. ${\cal L}_{mag}$ contains the baryon (here nucleons) and magnetic field interactions  through the electromagnetic field strength tensor ($F^{\mu\nu}$) and magnetic vector potential ($A^{\mu}$), as given in refs. \cite{dmeson_mag,bmeson_mag,charmonium_mag, Brodrik, Prakash, Wei}.

The magnetized Dirac sea contributes to the scalar densities of the nucleons, within the chiral model framework. The coupled equations of motion for the scalar fields are derived from the chiral model Lagrangian density under the classical scalar fields approximation, including the medium effects through the number ($\rho_{p,n}$) and scalar ($\rho^s_{p,n}$) densities of protons and neutrons in the nuclear matter. There are contributions of the Landau energy levels of protons and anomalous magnetic moments (AMMs) of nucleons through $\rho_{p,n}$ and $\rho^s_{p,n}$ due to the presence of ${\cal L}_{mag}$ term in the general Lagrangian density (\ref{genlag}), apart from the magnetized Dirac sea contribution described later. 
%\begin{equation}
 %   \mathcal{L}_{mag} = -\bar{\Psi}_iq_i\gamma_{\mu}A^{\mu}\Psi_i - \frac{1}{4}\kappa_i\mu_N\bar{\Psi}_i\sigma^{\mu\nu}F_{\mu\nu}\Psi_i - \frac{1}{4}F^{\mu\nu}F_{\mu\nu}
%\end{equation}
 
The masses of the baryons are generated by their interactions 
with the scalar mesons. 
The mass of baryon of species $i$
($i=p,n$ in the present work of nuclear matter)
is given as
\begin{equation}
M_i=-g_{\sigma i}\sigma -g_{\zeta i}\zeta -g_{\delta i}\delta. 
\label{mass}
\end{equation}
In the generalized SU(4) version of the chiral SU(3) model the interaction term involving the pseudoscalar open charm  $D$ and $\bar D$ mesons, \cite{dmeson_mag, amarindamprc, amarvdmesonTprc, amarvepja}
\begin{multline}
    \mathcal{L}_{DN} = -\frac{i}{8f_D^2}[3(\bar{N}\gamma^{\mu}N)(\bar{D}(\partial_{\mu}D)-(\partial_{\mu}\bar{D})D) + (\bar{N}\gamma^{\mu}\tau^aN)(\bar{D}\tau^a(\partial_\mu D)-(\partial_{\mu}\bar{D})\tau^a D)]  \\ 
    + \frac{m_D^2}{2f_D}[(\sigma + \sqrt{2}\zeta)(\bar{D}D) + \delta^a(\bar{D}\tau^a D)]  
-\frac{1}{f_D}[(\sigma + \sqrt{2}\zeta )(\partial_{\mu}\bar{D})(\partial^{\mu}D) + (\partial_{\mu}\bar{D})\tau^a (\partial^{\mu}D)\delta^a]  \\
+ \frac{d_1}{2f_D^2}(\bar{N}N)(\partial_{\mu}\bar{D})(\partial^{\mu}D) + 
 \frac{d_2}{4f_D^2} (\bar{N}N)(\partial_{\mu}\bar{D})(\partial^{\mu}D) + (\bar{N}\tau^a N)[(\partial_{\mu}\bar{D})\tau^a (\partial^{\mu}D)]
 \label{DN}
    \end{multline}

    In equation (\ref{DN}), we have the $D$ $(D^0,D^+)$ and $\bar{D}$ $(\bar{D}^0,D^-)$ mesons doublet. The (first) Weinberg-Tomozawa term is attractive for the $D$ mesons, repulsive for the $\bar{D}$ mesons. The (second) scalar meson exchange term is attractive for the $D$ and $\bar{D}$ mesons. The remaining terms are the range terms in the chiral model. The parameters $d_1$ and $d_2$ are determined by fitting the empirical values of kaon-nucleon scattering lengths for $I=0$ and $I=1$ channels.

The sea contribution for the spin-1/2, electrically charged fermions is given by the free energy of the magnetized vacuum \cite{haber} 
\begin{equation}
    \Omega_{N,sea} = -\frac{|qB|}{2\pi}\sum_{\nu=0}^{\infty}\alpha_{\nu}\int_{-\infty}^{\infty}\frac{dk_z}{2\pi}\epsilon_{k,\nu}.
    \label{ds}
\end{equation}
where, $\nu$ denotes the Landau level and sum over this quantity is implied. The single particle energies are $\epsilon_{k,\nu} = \sqrt{k_z^2+2\nu|qB|+M^2}$ when AMM of the fermion is taken to be zero, and $\alpha_{\nu}(=2-\delta_{\nu 0})$ is the spin degeneracy of each Landau level which is 1 for the lowest level with $\nu=0$. For non-zero AMM of the charged fermions, the single fermion energies turn into $\epsilon_{k,\nu, s} = \sqrt{k_z^2+(\sqrt{2\nu|qB|+M^2}+s\kappa_f B)^2}$, with an extra summation over the spin-projection $s=\pm1$ in eq.(\ref{ds}) with $\alpha_{\nu}=1$ for all $\nu$. In the mean-field approximation, $ \Omega_{N,sea}$ takes the form of free fermions by including all interactions to the in-medium mass term as in eq.(\ref{mass}), is not a pure vacuum term. Equation (\ref{ds}) involves divergent integral. The various regularization schemes of dimensional regularization or, proper time method have been employed in the literature to regularize the divergent integral \cite{haber, menezes, ferrar} and renormalizing the remaining term by proper scaling of the physical quantities like charge and magnetic field etc. \cite{haber}. The free energy of the Dirac sea contribution is minimized to give the scalar fields equations of motion, with the impact of the sea term on their solutions in magnetic field. 

The dispersion relation for the $D$ mesons are obtained from the Fourier transforms of the equations of motions of the mesons obtained from the chiral effective Lagrangian, given by

\begin{equation}
-\omega^2+ {\vec k}^2 + m_{D(\bar D)}^2
 -\Pi_{D(\bar{D} )}(\omega, |\vec k|)=0,
\label{dispddbar}
\end{equation}
where $\Pi_{D(\bar{D})}$ denotes the self energy and $m_{D(\bar{D})}$ is the vacuum mass
of the $D$ ($\bar{D}$) meson.
The self energies of $D$ meson ($D^0$, $D^+$) and $\bar D$ meson doublet (${\bar D}^0$, $D^-$) are given in terms of the scalar fields fluctuations,
the number and the scalar densities of the nucleons, as given in ref.\cite{dmeson_mag}.
For the charmed, strange meson $D_s$, the corresponding mass is 
%\begin{equation}
%-\omega^2+ {\vec k}^2 + m_{D_s}^2
% -\Pi_{D_s}(\omega, |\vec k|)=0,
%\label{dispdsdbar}
%\end{equation}
$m_{D_s}$, $\Pi_{D_s}$ is the $D_s$ meson self energy, which is same for $D_s^{+}$ and $D_s^{-}$ mesons in nuclear matter \cite{DP_AM_Ds}
\begin{eqnarray}
\Pi (\omega, |\vec k|)_{D_s} &= & \Big[\frac{d_1}{2f_{D_s}^2}(\rho_p^s + \rho_n^s  ) - 
\frac{\sqrt{2}}{f_{D_s}}(\zeta'+\zeta_c')\Big](\omega^2-\Vec{k}^2) + \frac{m_{D_s}^2}{\sqrt{2}f_{D_s}}(\zeta'+\zeta_c')
\label{selfds}
\end{eqnarray}
In equation (\ref{selfds}), $\zeta'(=\zeta-\zeta_0)$ is the fluctuation of the strange scalar-isoscalar field $\zeta (\sim \langle\bar{s}s\rangle)$. The value of $f_{D_s}$ is determined from the particle data group (PDG) as 250 MeV \cite{PDG}. The solution of equation (\ref{dispddbar}) for variable $\omega$ at $|\vec k|=0$ gives the in-medium mass of the corresponding pseudoscalar open charm meson. To obtain this, the coupled equations of motion of the scalar fields ($\sigma$, $\zeta$, $\delta$ and $\chi$), as derived from the chiral model Lagrangian by taking classical scalar fields assumption, are solved at given 
baryon density $\rho_B$, isospin asymmetry $\eta=(\rho_n-\rho_p)/(2\rho_B)$, and magnetic field $|eB|$. 

In the presence of an external magnetic field, for e.g. $\vec{B}$ along the $z$ direction, the energy level of an electrically charged point particle with spin $S$, mass m is given by $E_{n,S_z}(p_z)=\sqrt{p_z^2 +(2n+1+2S_z)|eB| +m^2 }$; with $p_z$ as the continuous momentum along the z-axis of $B\hat{z}$, $S_z=-S, -S+1, ..S$, is the spin-component along the magnetic field direction and $n$ is the Landau level \cite{gubler}. As mentioned in refs.\cite{lll1, lll2}, the minimal effective mass corresponding to the lowest energy state with $p_z=0$ can be considered as the ground state mass of the charged point-like particle. For spin-0, charged open charm mesons,
\begin{equation}
m^{eff}_{D^\pm}=\sqrt {{m^*_{D^\pm}}^2 +|eB|},
\label{mdpm_landau}
\end{equation}
%whereas for the neutral ($D^0$ and $\bar {D^0}$) mesons,
%the effective masses are given as
%\begin{equation}
%m^{eff}_{D^0 (\bar {D^0})}=m^*_{D^0 (\bar {D^0})}.
%\label{md0d0bar}
%\end{equation}
It is referred to as the point particle correction ignoring the internal structure of the mesons. However, in our work, the medium effects are incorporated mainly due to the interactions of the $D$ mesons with the nucleons and scalar mesons within the generalized chiral effective model. The contributions of Dirac sea and the Fermi sea in the magnetic matter are obtained in this way apart from the correction due to the point particle assumption. Although, at weak field region, higher Landau levels ($n\geq 0$) are situated infinitesimally close to the ground state $n=0$. In ref.\cite{gubler}, the sum over all Landau levels for the charged $D$ mesons have been taken into account using the formalism of QCD Borel sum rule. In equations (\ref{mdpm_landau}),
$m^*_{D^\pm}$ denote the masses of the corresponding $D$ mesons calculated from the solutions of eq.(\ref{dispddbar}) for $\omega$ at $|\vec k|=0$. For spin-1 mesons the ground state mass due to the point particle correction is therefore given by 
\begin{equation}
m^{eff}_{{D^*}^\pm}=\sqrt {{m^*}_{{D^*}^\pm}^2
+(2S_z+1)|eB|},
\label{mdpmstr_landau}
\end{equation}
 It is assumed that the mass shifts of the vector open charm 
($D^*$ and $\bar {D}^*$) mesons (which have the same quark-antiquark
constituents as $D$ and $\bar D$ mesons) are identical to the mass shifts of the pseudoscalar mesons $D$ and $\bar D$ mesons, calculated within the chiral effective model \cite{open_charm_mag_AM_SPM}.
This is in line with the QMC model where the masses of the hadrons are obtained from the modification of the scalar density of the light quark (antiquark) constituent of the hadron\cite{Hosaka}. 
The in-medium mass of the vector open charm mesons are thus given by \cite{upslndw_mag}
\begin{equation}
    m^*_{D^*(\bar {D}^*)}-m_{D^*(\bar {D}^*)}^{vac} 
= {m^*}_{D(\bar D)} - m_{D(\bar D)}^{vac},
\label{mdstrdbatstr}
\end{equation}
Using these values of $m^*_{D^*}$ in eq.(\ref{mdpmstr_landau}), the point particle corrections for the vector, charged $D^*$ mesons are obtained for the spin projections of $S_z=-1,0,1$ along $B\hat{z}$ direction. 

In an external magnetic field there is another important effect to be considered due to the mixing between the pseudoscalar (P) and the longitudinal component ($S_z$=0) of the vector ($V$) mesons. The PV mixing effect is observed to lead to an appreciable increase (decrease) in the mass of the longitudinal component of vector (pseudoscalar) mesons with increasing magnetic field \cite{cho113,gubler,cho91,suzuki,Alford,charmdw_mag,open_charm_mag_AM_SPM,upslndw_mag,Iwasaki}. In the present study, the PV mixing effect is considered for the neutral ($D^0-{{D}^{*0}}$, $\bar {D}^0-\bar {{D}}^{*0}$
) and charged ($D^{\pm}-{{D^*}^{\pm}}$) open charm mesons, accounting for the magnetized Dirac sea contribution on the masses. The PV mixing effect arise due to an effective hadronic Lagrangian accounting for the $PV\gamma$ interaction
vertices \cite{charmdw_mag, cho91,cho113, open_charm_mag_AM_SPM},
 \begin{equation}
     \mathcal{L}_{PV\gamma}=\frac{g_{PV}}{m_{av}}
e\tilde{F}_{\mu\nu}(\partial^{\mu} P)V^{\nu},
\label{pvg}
 \end{equation}
where P and $V^\mu$ represent the pseudoscalar and the vector meson fields, respectively, $\tilde{F}_{\mu\nu}$ is the dual electromagnetic field strength tensor and $m_{av}$ is the average of the masses of the pseudoscalar and vector mesons, $m_{av} = (m_P+m_V)/2$. The masses of the P and $V^{||}$ mesons due to the PV mixing effects can thus be written as \cite{cho91, charmdw_mag}
\begin{equation}
m^{2\ {(PV)}}_{P,V^{||}}=\frac{1}{2} \Bigg ( M_+^2 
+\frac{c_{PV}^2}{m_{av}^2} \mp 
\sqrt {M_-^4+\frac{2c_{PV}^2 M_+^2}{m_{av}^2} 
+\frac{c_{PV}^4}{m_{av}^4}} \Bigg).
\label{mpv2}
\end{equation} 
%\begin{equation}
  %    m_{V^{||},P}^2 = {m}_{V,P}^2 \pm \frac{c_{PV}^2}{{M_{-}}^2}
  %    \label{mpv2}
%\end{equation}
where, $M_{-}^2 = m_V^2 - m_P^2$ 
and $c_{PV}=g_{PV}|eB|$; with $m_{P,V}$ being the effective masses of the pseudoscalar and the vector mesons, given by equations (\ref{mdpm_landau}),
(\ref{mdpmstr_landau}),
as calculated in the magnetized nuclear matter within the chiral effective model.

The coupling parameter of the effective interaction Lagrangian or the PV mixing strength, $g_{PV}$ is fitted from the experimentally known radiative decay width as follows 
\begin{equation}
    \Gamma_{V\rightarrow P \gamma} = \frac{e^2}{12\pi} \frac{g_{PV}^2 p_{cm}^3}{m_{av}^2},
    \label{rad}
\end{equation}
where $p_{cm}=\frac{m_V^2-m_P^2}{2m_V}$ is the center of mass momentum of the final state particles. For the charged $D^{*\pm}$ mesons the measured radiative decay width of $\Gamma\ [D^{*\pm}\rightarrow D^{\pm} +\gamma]=1.33$ keV, is used to calculate the mixing strength between $D^{*\pm}$ and $D^{\pm}$ from eq.(\ref{rad}). 
The total decay width for the neutral $D^{*0}$ meson has not yet been measured, although the decay branching ratio of $\Gamma(D^{*0} \rightarrow D^0 \pi^0):\Gamma(D^{*0} \rightarrow D^0 \gamma) = 64.7:35.3$ is known \cite{PDG}. From isospin considerations, the decay constant for the $D^{*0} \rightarrow D^0 \pi^0$ can be obtained. Thus the decay width for the neutral $D^{*0} \rightarrow D^0 \pi^0$ can be determined. From the above mentioned decay branching ratio, the value for $g_{D^0D^{*0}}$ can be estimated %The mixing strength for the neutral mesons ($g_{D^0D^{*0}}$), involved in the radiative decay width can be calculated if the decay width  of  
 %$D^{*0}\rightarrow D^0 \pi^0$ mode is obtained. The decay constant for this neutral hadronic decay mode is related to that of the corresponding strong decay modes of the charged $D^{*\pm}$ mesons by the isospin symmetry considerations 
 \cite{gubler},
 \begin{equation}
     g[D^{*0} \rightarrow D^0 + \pi^0] = \frac{1}{\sqrt{2}}\ g[D^{*\pm} \rightarrow D^0 + \pi^{\pm}]
     \label{gg1}
 \end{equation}
 \begin{equation}
     g[D^{*0} \rightarrow D^0 + \pi^0] = g[D^{*\pm} \rightarrow D^{\pm} + \pi^0]
     \label{gg2}
 \end{equation}
 The hadronic effective Lagrangian of $\mathcal{L}_{D^*D\pi}= g\pi(\partial^{\mu}D)D^{*}_{\mu}$ is formulated to calculate the hadronic decay widths of $D^{*}\rightarrow D\pi$ modes from the imaginary part of the $D-\pi$ loop. The decay widths corresponding to the charged $D^{*\pm}$ mesons are fitted to their measured values \cite{PDG} to obtain the charged mesons decay constants. By equations (\ref{gg1})-(\ref{gg2}), the decay constant corresponding to $D^{*0}\rightarrow D^0 \pi^0$ mode is 
 obtained which is further used to calculate the decay width of $D^{*0}\rightarrow D^0 \pi^0$. Therefore, the decay branching ratio for the neutral mesons give the value of the PV mixing strength for the neutral mesons $g_{D^{*0}D^0}$. The effective Lagrangian $\mathcal{L}_{D^*D\pi}$  consists of only the leading order term in derivative, ignoring the higher derivative contributions which may also give impact to the calculated decay width. Due to this uncertainty, a method based on the constituent quark model have been proposed to determine the mixing strength between $D^{\pm}$ and $D^{*\pm}$ mesons \cite{gubler}, in which the $u$ and $d$ quarks are taken to be non-relativistic in the heavy charm quark limit. The mixing-strength between pseudoscalar and vector open charm mesons in the constituent quark model depends on the constituent quark masses and the quark charge, and is dominated by its light quark component. In ref.\cite{gubler}, the mixing strength corresponding to the neutral mesons is obtained from the ratio of $|\frac{g_{D^0D^{*0}}}{g_{D^{\pm}D^{*\pm}}}|=3.5$, to be 3.278 which is very close to our estimate of 3.44 determined using the previous method of effective Lagrangian. However, in the constituent quark model, the values of the constituent light quark $u,\ d$ masses can be changed with magnetic field as has been estimated in the context of NJL model in various quark matter studies at finite magnetic field \cite{ammc, menezes}. These modifications may lead to a change in the values of the mixing strengths in the effective Lagrangian eq.(\ref{pvg}), with the variation in magnetic field. In the present work, we rely on the framework of chiral effective Lagrangian to compute the magnetic field dependence of chiral order parameters, with the baryons and mesons as the effective degrees of freedom. To consider such effect on the change of constituent quark mass with magnetic field, is not within the scope of the present framework.

The radiative decay for $D_s^{*}$ has been observed but not measured experimentally. We use the Hamiltonian formulation to calculate the spin-mixing between spin-0 ($D_s$) and longitudinal component of spin-1 ($D_s^{*}$) mesons. In the presence of an external magnetic field, the Hamiltonian accounting for the spin-magnetic field interaction is given by \cite{upslndw_mag, Alford, Iwasaki}
\begin{equation}
    H_{spin-mix.} = -\sum_{i=1}^{2} \vec{\mu}_{i}.\vec{B} 
    \label{hamil}
\end{equation}
where, $\vec{\mu_i}=g'q_i \vec{S_i}/2m_i$ is the quark magnetic moment for the $i^{th}$ flavor present in the bound states of $D_s$ mesons (with constituent quark structure $\bar{q}_2q_1;\ q_{1,2}=s,c$). In this equation, $g'$ is taken to be 2. $q_i$ is the electric charge (in units of the electron charge $|e|$), $\vec{S_i}$ denotes the spin and $m_i$ is the mass of the $i^{th}$ quark flavor. The interaction Hamiltonian, (\ref{hamil}) leads to a rise (drop) in the mass of the longitudinal component of the vector (pseudoscalar) meson as follows
\begin{equation}
    m^{PV}_{V^{||},P} = m_{V,P} \pm \Delta m_{sB},
    \label{pv3}
\end{equation}
with $ \Delta m_{sB} = \frac{\Delta E}{2} ((1+x^2)^{1/2}-1)$;  $x = \frac{2}{\Delta E}\frac{(-g'|eB|)}{4}(\frac{q_1}{m_1}-\frac{q_2}{m_2})$; $\Delta E = m_V - m_P$, is the mass difference between the vector and pseudoscalar ($D^{*}_s$ and $D_s$) mesons calculated within the chiral effective model. The effective masses due to spin-mixing are obtained from eq.(\ref{pv3}), accounting for the DS effects in the magnetized nuclear matter.
 
 The mass shifts of the charmonium states have been evaluated from the modifications of the gluon condensate in nuclear medium by the QCD second order Stark effect, in the limit of large charm quark mass \cite{Lee}. In \cite{charmonium_mag, amarvdmesonTprc, amarvepja}, the mass shift is proportional to the medium modifications of the scalar gluon condensate $\langle\frac{\alpha_s}{\pi}G_{\mu\nu}^aG^{a\mu\nu}\rangle$, which is simulated by a scale-invariance breaking logarithmic potential in the scalar dilaton field $\chi$ within the chiral effective model. In-medium effects of density, isospin asymmetry, and magnetic field, are incorporated through the number and scalar densities of nucleons in the solutions of the scalar fields $\sigma,\ \zeta,\ \delta$ and $\chi$. As described in section \ref{sec2}, the additional effects of the magnetized Dirac sea are investigated in the present study and is considered to calculate the in-medium partial decay widths of charmonium going to open charm mesons, which may give rise to important phenomenological impacts in the production of these mesons as well as in the suppression of $J/\psi$. In chiral model, the scalar gluon condensate is expressed in terms of the fourth power of the scalar dilaton field in the limit of zero current quark masses. Thus introducing the mass shifts from the medium modifications of the scalar dilaton field $\chi$, in the magnetized nuclear matter including the Dirac sea contribution, as 
\begin{equation}
    \Delta m_{\Psi} = \frac{4}{81}(1-d)  \int dk^2 \Bigg \langle \Bigg | \frac{\partial \psi(\Vec{k})}{\partial \Vec{k}}\Bigg |^2\Bigg \rangle \frac{k}{k^2/m_c + \epsilon}(\chi^4-\chi_0^4)
    \label{psi3770}
\end{equation}
with 
\begin{equation}
   \Bigg \langle \Bigg| \frac{\partial \psi(\Vec{k})}{\partial \Vec{k}}\Bigg|^2\Bigg\rangle = \frac{1}{4\pi} \int \Bigg| \frac{\partial \psi(\Vec{k})}{\partial \Vec{k}}\Bigg|^2 d\Omega
\end{equation}
Here $m_c$ is the charm quark mass of 1.95 GeV \cite{charmonium_mag}. $m_\Psi$ is the vacuum mass of $\Psi(3770)$, $\epsilon = 2m_c - m_\Psi$ is its binding energy. The bound state wave function of $\Psi(3770)$ is obtained using the harmonic oscillator potential model to calculate the mass shifts. The partial decay widths of $\Psi(3770) \rightarrow D\bar{D}$ are calculated in the $^3P_0$ model \cite{Friman, amarvepja}. In the harmonic potential, the wave function of $\Psi(1D)$ state is given for $N=1$ and $l=2$, as \cite{Friman}
\begin{equation}
    \psi_{N,\ l}(r) = N' \times Y_l^m(\theta,\phi)(\beta^2r^2)^{l/2}e^{-\frac{\beta^2r^2}{2}}L^{l+1/2}_{N-1}(\beta^2r^2).
    \label{wavef}
\end{equation}
 $N'$ is the normalization constant. $L^k_p(z)$ is the associated Laguerre polynomial. $\beta^2 = M\omega/\hbar$ characterizes the strength of the harmonic potential. $M=m_c /2$ is the reduced mass of charm quark-antiquark bound state. In eq.(\ref{psi3770}), $\psi(\Vec{k})$ is the Fourier transform of the wave function in coordinate space, eq.(\ref{wavef}) for $N=1$ and $l=2$. Thus, the Gaussian function multiplied by a polynomial is generated using the harmonic potential in the bound state problem of charm quark and antiquark, and used in the context of $^3P_0$ model \cite{Friman, amarvepja} which considers the internal structure of mesons as a quark-antiquark bound state to calculate the decay widths. The value of $\beta$ for $\Psi(3770)$ is taken to be 0.37 GeV by fitting from its root mean squared radius $\langle r^2 \rangle=1\ fm^{2}$ \cite{Lee}. The sizes of the charmonium states are  related to the strength of the harmonic potential $\beta$. Variation of $\beta$, hence of $r_{rms}$ with $|eB|$ can be obtained from the mass shifts of the corresponding state, for $\Psi(3770)$ it is given by $\beta \delta\beta=\frac{M}{7} \delta M_{\Psi(1D)}$ \cite{Friman, amarvepja}. The r.m.s radius of $\Psi(3770)$ as a function of magnetic field is plotted in section (\ref{sec4}), from its mass shifts. The variation of 
r.m.s radii for the different states of charmonium and bottomonium have been studied in hadronic matter \cite{AM_DP_upsilon,amarvepja, Friman}.  

\section{The $^3P_0$  model}

\label{sec3}
In this section we discuss the in-medium partial decay widths of ($D^* \rightarrow D  \pi$, $\bar{D}^* \rightarrow \bar{D}  \pi$), 
%$D_s^{*\pm} \rightarrow D_s^{\pm}  \pi^0$ ) 
and $(\Psi(3770)\rightarrow D\bar{D})$, by taking into account the internal structures of the parent and the outgoing mesons using the $^3P_0$ model \cite{Friman}.
\subsection{Decay of the Open Charm mesons}
\label{subsecA} 
The various in-medium effects discussed in the previous section on the masses of the open charm mesons $(D,D^*)$ in a magnetized nuclear matter, are considered in the decay width calculation. The in-medium masses result in the modified decay widths for the various channels. From the $^3P_0$ model, the decay width of the vector open charm meson $D^*$ decays into a $D$ and $\pi$ mesons is given by \cite{charmdecay_mag, 3p0f, Friman}

\begin{equation}
     \Gamma_{D^*\rightarrow D\pi} =  {\frac{\sqrt{\pi}E_DE_\pi\gamma^22^8r^3(1+r^2)^2x^3}{2m_{D^*}\ 3(1+2r^2)^5}}\ exp\Bigg(-\frac{x^2}{2(1+2r^2)}\Bigg)
\label{dw1}     
\end{equation}
In this expression, $m_{D^*}$ is the mass of the parent $D^*$ meson. $E_D$ and $E_\pi$ are the energies of the outgoing $D$ and $\pi$ mesons, respectively,
     \begin{equation}
         E_D = (p^2 + m_D^2)^{1/2} ;  \  E_\pi = (p^2 + m_{\pi}^2)^{1/2}
     \end{equation}
with $m_D$, the in-medium mass of $D$ meson and $m_\pi$ is the mass of $\pi$ meson. $p$ is the magnitude of the 3-momentum in the center of mass (c.o.m) frame.
\begin{equation}
    p = \bigg[\frac{m_{D^*}^2}{4} - \frac{m_D^2 + m_{\pi}^2}{2} - \frac{(m_D^2 - m_{\pi}^2)^2}{4m_{D^*}^2}\bigg]^{1/2}
\end{equation}
In equation (\ref{dw1}), $\gamma$ is the coupling constant related to the strength of the $^3P_0$ vertex \cite{3p0f,3p0g}. It signifies the probability for creating a light quark-antiquark pair. In the context of $^3P_0$ model, the wave functions of the $(D^*,D)$ mesons are obtained with the harmonic oscillator potential. The ratio $r = \beta/\beta_{avg}$, where $\beta$ is the strength of the harmonic potential of the parent $D^*$ meson state and $\beta_{avg}$ is the average strength of harmonic oscillator potential of the daughter $(D-\pi)$ mesons. The scaled momentum $x$ is defined as $x = p/\beta_{avg}$. The value of $\beta_D=0.31$ GeV, is consistent with the decay widths of $\psi(4040)$ to $D\bar{D}, D^*\bar{D}, D\bar{D^*},$ and $ D^*\bar{D^*}$ in vacuum \cite{3p1,amarvepja,Lee}. Decay widths depend on the variable, $x$, which is the center of mass momentum, $p$
in units of $\beta_{avg}$, as a polynomial multiplied by an exponential term. The in-medium partial decay widths of the charmed, strange mesons, $D_s^{*\pm}\rightarrow D_s^{\pm}\pi^0$ can be determined on the same footing as $D^*$ mesons, from eq.(\ref{dw1}) by using the corresponding masses of $D_s$ and $D_s^*$ mesons, with the coupling parameter and the harmonic oscillator strengths taken to be the same as the $D^*$ mesons decays. 
\subsection{Decay of $\Psi(3770)$ meson to $D\bar{D}$}
\label{subsecB} 
The hadronic decay width for the charmonium state, $\Psi(3770)$ going to $D\bar{D}$ mesons is calculated in this subsection, using the $^3P_0$ model. The in-medium masses of the parent and daughter particles are obtained within the chiral effective model, accounting for the effects of Dirac sea in the magnetized nuclear matter. The decay width is given by \cite{charmdecay_mag, Friman} 
\begin{multline}
\Gamma_{\Psi(3770) \rightarrow D\bar{D}} = \frac{ \gamma_{\Psi}^2\sqrt{\pi} E_D  E_{\bar{D}} }{2m_{\Psi (3770)}}\frac{2^{11} 5}{3^2}  \left(\frac{r_{\Psi}}{1+2r_{\Psi}^2}\right)^7 \times\\ x_{\Psi}^3 \left(1-\frac{1+r_{\Psi}^2}{5(1+2r_{\Psi}^2)}x_{\Psi}^2 \right)^2
 exp \left(-\frac{x_{\Psi}^2}{2(1+2r_{\Psi}^2)}\right)
 \label{dw2}
\end{multline}
with $E_D = (p_D^2 + m_D^2)^{1/2}$ and     $ E_{\bar{D}} = (p_D^2 + m_{\bar{D}}^2)^{1/2}$ and $ p_D = (m_{\Psi}^2/4 - m_D^2 )^{1/2}$. Similar to the previous decay channel $(D^* \rightarrow D \pi)$,  $\gamma_{\Psi}$ signifies the probability for creating the light quark-antiquark pair, is chosen to be 0.33 \cite{amarvepja}, thus reproducing the observed decay widths of $\Psi(3770)$ to $D^+D^-$ and $D^0\bar{D}^0$ in vacuum. The ratio $r_{\Psi}=\beta_{\Psi} / \beta_D $ with $\beta_{\Psi} = 0.37$ and $\beta_D = 0.31$ (in GeV) are obtained from the mean squared radius of $\Psi(3770)$ and D meson states. %$\beta_{D} = 0.31$ GeV, is consistent with decay widths of $\Psi(4040)$ to $D\bar{D}, D^*\bar{D}, D\bar{D}^*$ and $D^*\bar{D}^*$ in vacuum \cite{Lee}. 
In (\ref{dw2}), $x_{\Psi}=p_D/ \beta_D$ is the scaled momentum.

\section{Results and Discussions}
\label{sec4}

In the present work, we study the in-medium properties of masses and decay widths of the open charm and $\Psi(3770)$ mesons in  magnetized, asymmetric nuclear matter, accounting for the Dirac sea effects. The in-medium masses of the pseudoscalar ($D$, $\bar D$ and $D_s^{\pm}$) and vector ($D^*$, $\bar{D}^*$ and $D_s^{*\pm}$) open charm mesons are investigated within the generalized version of the chiral effective model. The masses thus obtained are used to find the in-medium partial decay widths of the $D^*\rightarrow D\pi$ ($\bar{D}^*\rightarrow \bar{D}\pi$) and $\Psi(3770) \rightarrow D\bar{D}$ channels. The in-medium masses of the pseudoscalar $D$ ($\bar D$) and $D_s^{\pm}$ mesons are obtained due to their interactions with the nucleons and the scalar mesons in the effective chiral model. The effects of an external magnetic field are taken into account through both the Dirac sea (denoted as DS) and the Fermi sea, i.e., the Landau quantization of protons and anomalous magnetic moments (AMM) of the nucleons from the nuclear matter part. 

In our present study, the Dirac sea contribution is accounted for by the renormalized free energy of the magnetized vacuum which itself is a part of the total nucleonic free energy. Its application to nuclear matter is what we deal with in the context of chiral SU(3) model by minimizing the Dirac sea contribution of the free energy in the minimization of the thermodynamic potential of the chiral effective model. This is not a pure vacuum term in that sense as it includes the nucleon's mass term representing baryon-scalar meson interactions in the mean-field approximation. The contributions of the Landau energy levels of charged fermions, i.e., protons as well as their anomalous magnetic moment are studied in the computation of magnetized Dirac sea in this work. The regularization of the divergent integral can be done by means of proper time method or, dimensional regularization leading to similar results. The details of the calculation have been described in the literature in the context of quark matter and nuclear matter applications for zero AMM of the charged fermions \cite{menezes,haber}. However in line with the argument given in \cite{haber}, the correction due to the sea contribution in absence of magnetic field is very small as compared to the magnetized DS part, hence it is not considered here. The magnetic field-dependent sea part is responsible for the phenomenon of magnetic catalysis in vacuum and in specific case for nuclear matter, which is of concern in the present study. The field-independent sea contribution is of main concern in theory where it induces the chiral symmetry breaking effect in vacuum for coupling strength 
larger than a critical coupling for e.g., in the NJL model, which in our case is already incorporated by the nucleonic mass term through baryon-scalar meson interactions. Finally the renormalized magnetized sea is independent of any renormalization scale (either magnetic field or nucleon mass) after minimization of the free energy. In this work, the AMM of the charged fermions is considered with the contribution of lowest Landau level in the single particle energies. The single particle energies of charged fermions incorporate the summation over all Landau levels for zero AMM while calculating the renormalized magnetized sea contribution. For non-zero AMM, however to avoid the complexity of the renormalization of the sea term and given the fact that the lowest Landau level is a perfectly valid approximation in the strong field limit, we restrict our calculations to the lowest level contribution with non-zero AMM. Mentioned below are the exact no. of Landau energy levels contributing at different magnetic field strength in magnetized matter. We compare the solutions of the scalar isoscalar field $\sigma$ [$\sim (\langle \bar{u}u\rangle+\langle\bar{d}d\rangle$)], corresponding to the case of with AMM, lowest Landau level and the other case of without AMM, sum over all Landau levels in the calculation of magnetized Dirac sea effects leading to the phenomenon of magnetic catalysis in the magnetized vacuum.  

\begin{figure}
    \centering
    \includegraphics[width=9cm]{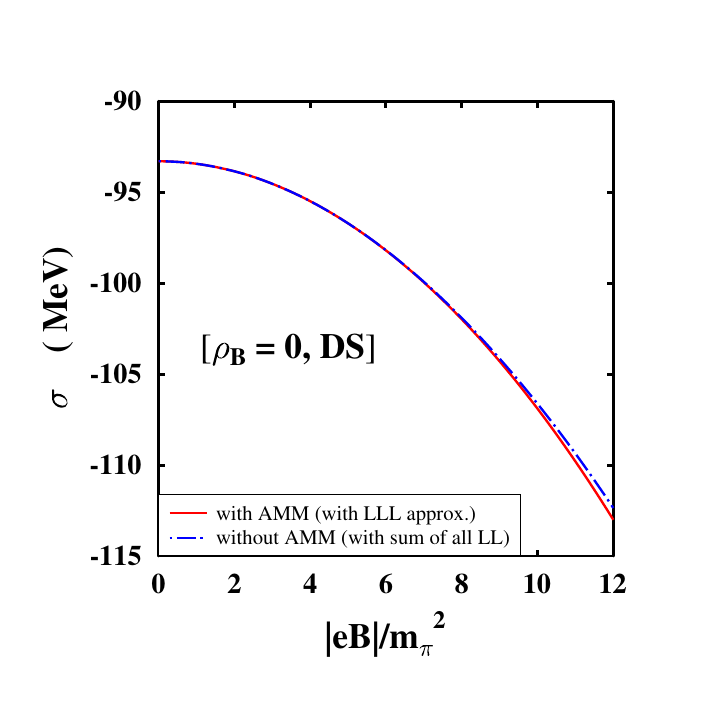}
\vspace{-1.2cm}
     \caption{The scalar isoscalar field $\sigma$ (in MeV) is plotted as a function of $|eB|/m_\pi^2$, accounting for the magnetized Dirac sea (DS) in vacuum $\rho_B$=0. Two cases are compared- (i) sum over all Landau levels at zero AMM, and (ii) only lowest Landau level approximation at non-zero AMM, in the single-fermion energies of charged fermions.} 
    \label{sigm}
\end{figure}

We have considered the effects of both spin (up/down) projections of protons and neutrons when accounting for their AMM in the magnetized Fermi sea and determined the required no. of Landau levels contributing at different values of $|eB|$ at the nuclear matter saturation density $\rho_0$, at zero temperature. At $\rho_0$, $|eB|=1 m_{\pi}^2$, it is $\nu^{max}_{up}=2$ and $\nu^{max}_{down}=1$, at $|eB|=2 m_{\pi}^2$, $\nu^{max}_{up}=0$ and $\nu^{max}_{down}=1$, corresponding to up and down spin-projections of protons along the field axis. There is only $\nu=0$ level contribution for $|eB| > 2 m_{\pi}^2$. Hence, the LLL approximation at finite AMM in the magnetized Dirac sea calculation can be adopted at finite density as well as in vacuum. It is shown in figure (\ref{sigm}) by the variation of $\sigma$ with $|eB|$, that the solutions corresponding to (with AMM+LLL) and (without AMM+sum over all Landau levels)   coincide in the lower magnetic field region. Hence, our approach is justified in the low-field regime by the similarity between the solutions of the two cases.
%we have abandoned any weak-field approximation of fermionic propagator in the calculation of magnetized Dirac sea which was contradictory to the LLL point particle correction for the charged $D$ meson masses.  

As it has been illustrated in section (\ref{sec2}), the in-medium masses of the pseudoscalar open charm mesons are obtained through the interactions of $D$ mesons with the nucleons [$N=(p,n)$] and scalar mesons $\sigma$, $\delta$ ($\zeta$ for $D_s$ mesons) within the generalized version of chiral SU(3) model to SU(4). The medium effects of density, magnetic field are included in terms of the scalar fields fluctuations $\sigma'$, $\delta'$ ($\zeta'$ for $D_s$), number ($\rho_{p,n}$) and scalar densities ($\rho^s_{p,n}$) of nucleons to the masses via solutions of the dispersion relation eq.(\ref{dispddbar}). The thermodynamic potential of the chiral SU(3)$_L\ \times$ SU(3)$_R$ model is minimized with respect to the scalar, isoscalar fields $\sigma$, $\zeta$, isovector $\delta$, and the scalar dilaton field $\chi$ by incorporating the effects of density, isospin asymmetry and magnetic field through the scalar and number densities of protons and neutrons. The effect of magnetic field on the nuclear matter is accounted for by the Landau energy levels of protons and anomalous magnetic moments of the nucleons. The free energy of the magnetized vacuum is also minimized with respect to the scalar fields to obtain the coupled equations of motion of the scalar fields $\sigma$, $\zeta$, $\delta$ and $\chi$. The effects of anomalous magnetic moments of the quarks have been studied extensively in the context of Nambu-Jona-Lasinio (NJL) model while solving the meson masses in the constituent quark model as compared to the current model which treat the mesons and baryons as the effective degrees of freedom and any medium effects have been incorporated through various interaction terms of mesons and baryons. Whereas the AMMs of the light $u,\ d$ quarks have been taken into account in the $SU(2)$ NJL model to study its effect on the chiral transition at zero and finite temperature \cite{AMM}, the effect of charm quark's AMM is not well studied. In the effective mass formula given by eq.(\ref{mdpmstr_landau}), the (charged) mesons are treated as point-like particles by ignoring its internal quark structure, hence the effects of the constituent quarks AMMs for e.g. of $d,\ c$ are not considered within the scope of this formula. However, in the chiral effective model, while considering the meson-nucleon interactions of the pseudoscalar open charm mesons, the effects of nucleons' anomalous magnetic moments are incorporated in the magnetized nuclear matter in the Fermi sea apart from the AMMs of charged fermions in the magnetized Dirac sea contribution. 

In the magnetized vacuum, the summation over all Landau levels in the single fermion energies have been executed in the calculation of renormalized DS effects for zero AMM of the charged fermion. There is no approximation taken in this case since the Landau levels are very closely spaced at weak magnetic fields, and separating the lowest one would throw out important effects from the magnetic fields. In finite AMM calculation, for simplicity of the renormalization of magnetized vacuum, only LLL contribution is taken and the solutions obtained in this case match with the ones obtained for zero AMM in the weak-field region. 
At finite density we have shown the contributions of the maximum no. of Landau levels and it is reported that after $2 m_{\pi}^2$, there is only LLL effect in the magnetized Fermi sea calculation, hence taking the LLL approximation in this case for DS calculation with AMM is well justified as compared to the vacuum case. In the vacuum, we rely on the comparison of the scalar fields solutions and justify the use of LLL with AMM in magnetized vacuum.

\begin{figure}[h!]
    \centering
    \includegraphics[width=15cm]{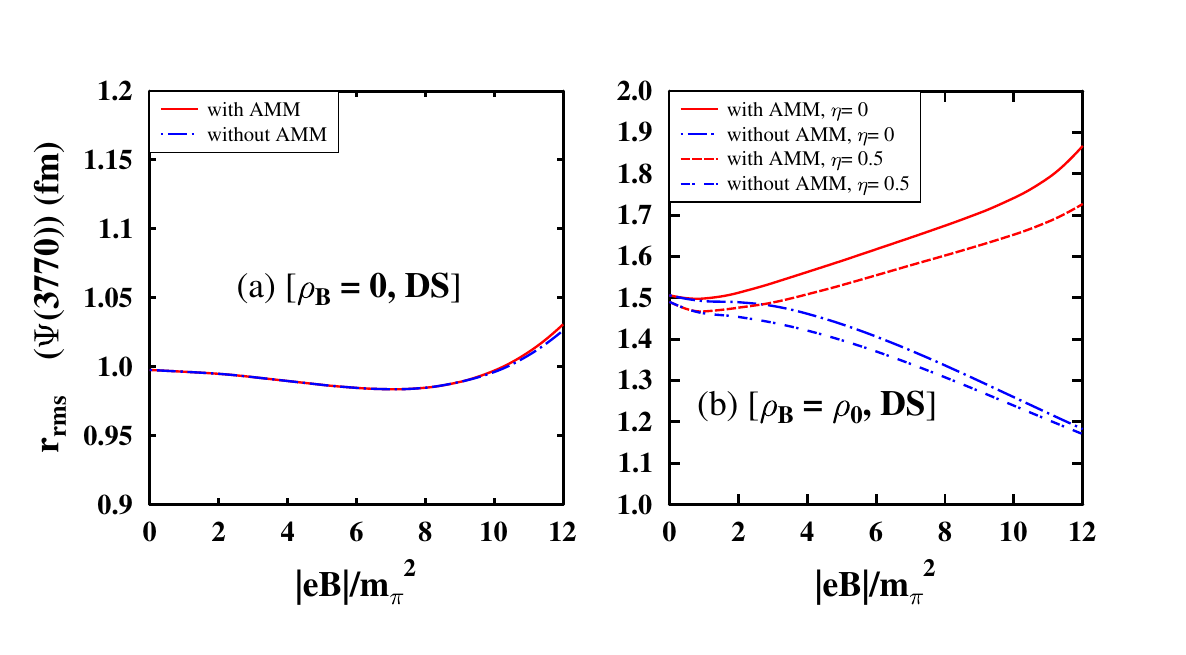}
\vspace{-0.8cm}
     \caption{The root mean square (rms) radius (in fm) of $\Psi(3770)$ state as a function of $|eB|/m_{\pi}^2$ including the Dirac sea (DS) effect, at $\rho_B = 0,\ \rho_0$ and $\eta=0,\ 0.5$, due to the medium modifications of the masses of the state. }
    \label{fig16}
\end{figure}

The variation of r.m.s size of the $1D$ state of charmonium, $\Psi(3770)$ with magnetic field is plotted in figure (\ref{fig16}) due to the modifications of its mass with $|eB|$, accounting for the magnetized Dirac sea effects.

In \cite{yoshida, Alford, yoshida1}, the two-particle Hamiltonian of the constituent quark model in a magnetic field has been constructed in the non-relativistic approach. The center-of-mass momentum of the system is not conserved due to the breaking of translational invariance by vector potential, instead one obtains a conserved quantity called pseudomomentum $\bf K$. Particular choice of the potential, either harmonic or, Cornell potential form is considered in solving the eigenvalue equation for the relative Hamiltonian of the particle-antiparticle pair, with the corresponding wave functions and energy eigenvalues. Solving the Schr$\ddot o$dinger equation for such system of particles with non-zero ${\bf K}$ lead to the quantization of the energy eigenvalues. In some constituent quark model studies at finite magnetic field involving neutral heavy-light mesons and states of heavy-quarkonia, the pseudomomentum is also taken to be zero to solve the eigenvalue problem for mesons at rest \cite{yoshida, yoshida1}. In this case the Hamiltonian preserves the rotational symmetry on the transverse plane of the magnetic field direction, as well as the reflection symmetry along the field direction. In our study of open charm mesons in magnetic nuclear matter within the generalized chiral effective model, the energy dispersion relation eq.(\ref{dispddbar}) is solved for zero three momentum of the pseudoscalar open charm mesons. The Landau energy level contributions for the charged open charm mesons are considered by taking only the lowest Landau level corresponding to its point particle correction to the ground state energy \cite{lll1,lll2}. The quantization of the energy eigenvalues for non-zero ${\bf K}$, as expressed in \cite{Alford} corresponding to harmonic potential for a bound state of quark-antiquark pair, for e.g., $J/\psi$, $\eta_c$, involve the contributions of magnetic field through the minimal substitution (${\bf p}\rightarrow ({\bf p}-e {\bf A})$) and spin-magnetic field interaction. The spin-magnetic field interaction for the system of particle-antiparticle pair lead to the mixing between the spin-0 (pseudoscalar) and longitudinal component of spin-1 (vector) mesons \cite{Alford, yoshida, yoshida1} as well as non-zero shift in the energy of the transverse components (only for heavy-light system) \cite{yoshida, Machado}. In the present work, the mixing between $\Psi(3770)^{||}-\eta_c(2S)$ as well as $D^{*||}-D$ meson states is studied from an effective hadronic Lagrangian involving $PV\gamma$ interaction vertices resulting in the mixing or level repulsion between the longitudinal component of vector and the pseudoscalar mesons in presence of an external magnetic field, which for the $D_s^{*||}-D_s$ mesons obtained from the coupling of spin with magnetic field. The effects of magnetic field on the hadron properties in our case are considered through the modifications of the QCD vacuum properties of light quark and gluon condensates in terms of the scalar fields within the framework of chiral effective model, in contrast to the non-relativistic approach of constituent quark model to study a meson system in an external magnetic field, which for dense matter case is not appropriate. For the heavy quarkonium states, the medium modifications are obtained through that of the scalar gluon condensate, in comparison to the approach of the non-relativistic constituent quark model, and hence do not consider the quantization of pseudomomentum but is consistent with the assumption of zero ${\bf K}$ for mesons at rest. Deriving the in-medium properties of mesons through the medium modifications of the QCD vacuum gives rise to comparable mass modifications with other contemporary studies of effective theory and non-relativistic quark model.     

At $\rho_B=0$, for non zero and zero AMMs of nucleons, the Dirac sea contributions lead to the increasing magnitudes of the scalar fields, $\sigma (\sim (\langle \bar u u\rangle + \langle \bar d d\rangle))$ and $\zeta (\sim \langle \bar s s\rangle)$ with magnetic field, i.e., the enhancement of the light quark condensates, an effect called magnetic catalysis (MC). At $\rho_B=\rho_0$, the magnitudes of the scalar fields tend to decrease with increasing magnetic field ($|eB|$) for non zero AMMs of the nucleons, leading to the decrement of the light quark condensates with $|eB|$, an effect called inverse magnetic catalysis (IMC), for symmetric ($\eta=0$) as well as asymmetric ($\eta=0.5$) nuclear matter. The opposite behavior is obtained, i.e., the light quark condensates increase with the magnetic field, if AMM is taken to be zero, indicating magnetic catalysis. The Dirac sea effects are realized due to the free energy density of magnetized vacuum, as discussed earlier. Weak-field approximation is not considered to incorporate the Dirac sea effects in the current analysis. For zero baryon density, magnetic field has impact on the QCD vacuum properties due to the magnetized Dirac sea, with no Landau level contribution of protons, which is only present at finite baryonic matter. 

The masses of the $D$ mesons undergo modification as per the generalized chiral effective model. Effects of densities and strong magnetic fields are incorporated in the values of the scalar fields which in turn depend on the Fermi sea (matter contribution), the Dirac sea (vacuum contribution). The in-medium masses of the pseudoscalar open charm mesons are obtained by solving the dispersion relations, equations (\ref{dispddbar}), for energy $\omega$ at $|\vec{k}|=0$, with their self energy functions given in terms of the number and scalar densities of nucleons and the scalar fields fluctuations $\sigma'$, $\zeta'_c$ and $\delta'$ for the $D(\bar{D})$ mesons and $\zeta'$, $\zeta'_c$ for $D_s^{\pm}$ mesons. However, the fluctuation in the heavy charm quark condensate $\zeta'_c$, is neglected in our present calculation \cite{dmeson_mag}. The in-medium masses of the vector open charm and charmed, strange mesons, $D^*(\bar{D}^{*})$ and $D_s^{*\pm}$ are obtained by calculating their mass shifts from the mass shifts of the corresponding pseudoscalar mesons using equation (\ref{mdstrdbatstr}). The additional effects of magnetic field due to the particle-like correction to the ground state mass, are considered on the masses of the electrically charged mesons, as given by equation (\ref{mdpm_landau}) for the pseudoscalar meson ($D^{\pm}$, $D_s^{\pm}$), and by (\ref{mdpmstr_landau}) for the vector meson ($D^{*\pm}$, $D_s^{*\pm}$) states. For the neutral mesons $D^0(\bar{D}^0)$ and $D^{*0}(\bar{D}^{*0})$, there is no such correction. An important contribution of magnetic field is obtained due to the interaction of the pseudoscalar and the longitudinal component of the vector mesons, which lead to a level repulsion between their masses. An effective hadronic Lagrangian (\ref{pvg}) is employed to study this effect for the open charm and charmonium states. The effective masses are thus calculated for the pseudoscalar and longitudinal component of the vector mesons from equation (\ref{mpv2}), accounting for the mixing between the ($D-D^{*}$) and ($\bar{D}-\bar{D}^{*}$) mesons. A Hamiltonian approach is adopted to find the spin-magnetic field interaction between ($D_s-D_s^{*}$) mesons, the effective masses for $D_s^{\pm}$ and $D_s^{*\pm||}$ are given by equation (\ref{pv3}). The in-medium partial decay widths of $\Psi(3770)\rightarrow D\bar{D}$ in terms of the medium modified parent and daughter particles' masses are studied in the present work accounting for the Dirac sea effects. The mass shift of $\Psi(3770)$ is calculated from the medium modified scalar dilaton field, $\chi$, within the chiral model framework, as given by equation (\ref{psi3770}). The coupling parameters $g_{PV}$ in equation (\ref{pvg}) are fitted from the observed radiative decay widths of the charmonium and open charm mesons. The effective mass for $\Psi(3770)$ state due to the PV mixing of $(\eta_c(2S)-\Psi(3770))$ is also calculated in the magnetized nuclear matter, accounting for the Dirac sea effects on their masses. 

The in-medium masses of the open charm and charmonium states as illustrated above, lead to the modifications of the hadronic decay widths of $D^*\rightarrow D\pi$ ($D^{*+}\rightarrow D^{+}\pi^0$, \ $D^{*+}\rightarrow D^0\pi^+$, $D^{*0}\rightarrow D^0\pi^0$); $\bar{D}^*\rightarrow \bar{D}\pi$ ($D^{*-}\rightarrow D^{-}\pi^0$, $D^{*-}\rightarrow \bar{D}^0\pi^-$, $\bar{D}^{*0}\rightarrow \bar{D}^0\pi^0$), $D_s^{*\pm}\rightarrow D_s^{\pm}\pi^0$ and $\Psi(3770) \rightarrow D\bar{D}$, accounting for the Dirac sea effects. A light quark-antiquark pair creation model or the $^3P_0$ model is used in our present investigation to study the hadronic decay widths for the vector open charm meson, $D^*$ going to two pseudoscalar mesons $D$ and $\pi$, using equation (\ref{dw1}). Due to the isospin considerations, the coupling constant $\gamma$, related to the $^3P_0$ vertex for the channels  $D^{*\pm}\rightarrow D^{\pm}\pi^0$, and $D^{*+}\rightarrow D^0\pi^+$ (with its charge conjugate mode, $D^{*-}\rightarrow \bar{D}^0\pi^-$) is chosen to be 0.265 and 0.368, respectively. The value of $\gamma$ for the $D_s^{*\pm}\rightarrow D_s^{\pm}\pi^0$ channel is taken to be the same as for the $D^{*\pm}\rightarrow D^{\pm}\pi^0$ channel. For $D^{*0}\rightarrow D^0\pi^0$ it is 0.264 (including the charge conjugate channel). The decay of $\Psi(3770)$ to $D\bar{D}$ can be possible by two decay modes, namely, $D^+D^-$ and $D^0\bar{D}^0$, which are calculated within the $^3P_0$ model to find the total decay width of $\Psi(3770)$, by equation (\ref{dw2}). 

%The PV mixing of ($D-{D^*}$ and $\bar D-\bar{D}^*$) states, result in a mass drop (rise) for the $D({D^*}^{||})$ as well as for the $\bar D(\bar {D}^{*||})$ mesons with increasing magnetic field. The masses of the charmed, strange mesons $D_s^{+}$ and $D_s^-$, are observed to be degenerate in the nuclear matter. In the self-energy functions of the open charm $D$ mesons \cite{dmeson_mag}, the vectorial Weinberg-Tomozawa interaction term is attractive for the $D$ mesons but repulsive for $\bar{D}$ mesons, the scalar meson exchange term is attractive for both $D$ and $\bar{D}$ and there are the range terms of $d_1$ and $d_2$. However, for the charmed, strange mesons in nuclear matter, as given by equation (\ref{selfds}), only a $d_1$ term exists and it contains contributions from the scalar densities of protons and neutrons. The Weinberg-Tomozawa and the range $d_2$ terms contain the contributions from the hyperon scalar densities and are absent in our present case, rendering similar behaviour to the $D_s^{+}$ and $D_s^{-}$ mesons in the nuclear medium. 
\begin{figure}
    \centering
    \includegraphics[width=15cm]{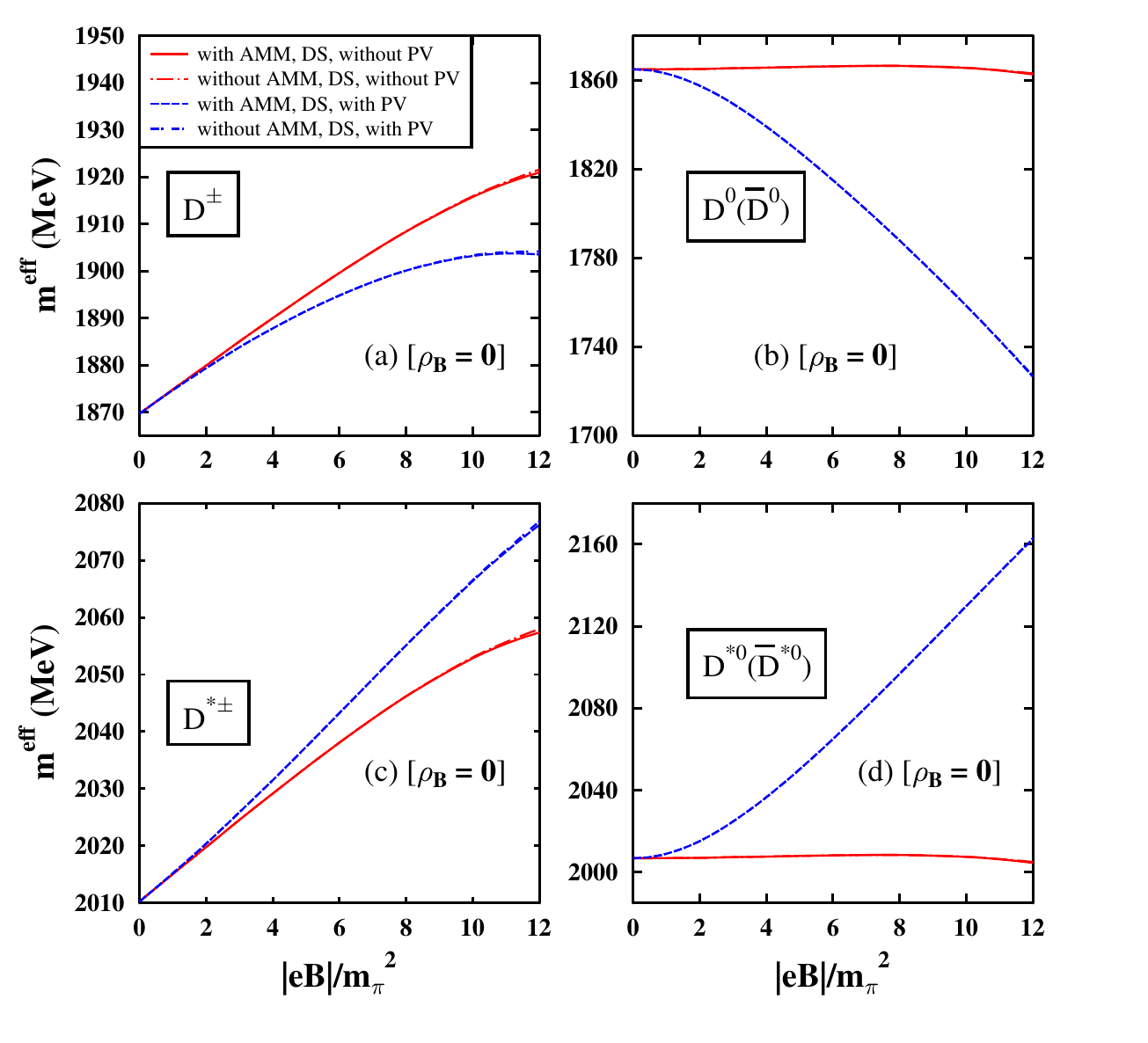}
\vspace{-0.8cm}
     \caption{The masses (in MeV) of $D^{\pm}$, $D^0$ ($\bar {D}^0$), $D^{*\pm}$ and ${D^*}^0$ ($\bar {{D}}^{*0}$) mesons are plotted as functions of $|eB|/m_\pi^2$, in (a), (b), (c) and (d) respectively, at $\rho_B$=0, including the Dirac sea (DS) effects, with and without the PV mixing between ($D^+-{{D^*}^{+||}}$, $D^--{{D^*}^{-||}}$, $D^0-{{D}^{*0||}}$, $\bar {D}^0-\bar {D}^{*0||}$) states. The point-particle correction for the charged meson is included at finite $|eB|$.} 
    \label{fig1}
\end{figure}

In figure \ref{fig1}, the masses of the charged as well as the neutral pseudoscalar and vector $D$ mesons are plotted with variation in magnetic field, $|eB|$, accounting for the Dirac sea as well as PV mixing effects at $\rho_B=0$. The effects of AMMs of the charged fermions in the Dirac sea are taken into account and compared to the case of zero AMM solutions.  There is only Dirac sea contribution at zero density (with no matter component). The scalar fields ($\sigma$, $\zeta$, $\delta$ and $\chi$) solutions in this case are obtained due to the non-zero effects from the magnetized Dirac sea contribution to the potential, which leads to the increasing magnitudes of light quark condensates with magnetic field for both zero and non-zero AMMs of the charged fermions. This effect is known as magnetic catalysis. The effects of this phenomenon on the in-medium masses of the pseudoscalar and vector open charm mesons are shown with the additional effect due to the PV mixing between the corresponding P-V partners and point particle Landau level correction for the ground state energy of $D^{\pm}$, $D^{*\pm}$ mesons. The masses for $D^\pm$ are observed to be identical to each other, similarly for the neutral mesons sector ($D^0$ and $\bar{D}^0$). This can be understood from the self energy functions $\Pi_{D(\bar{D})}$ of $D(\bar{D})$ mesons in the dispersion relation (\ref{dispddbar}). For $\rho_B=0$ (hence $\rho_p$ and $\rho_n$ are both zero), the Weinberg-Tomozawa contribution turns out to be zero. At $\rho_B=0$, there is no difference in the mass of the open charm $D$ mesons due to the contribution of the AMM of the charged fermions. For the charged pseudoscalar $D^{\pm}$ mesons (\ref{fig1}a and \ref{fig1}b), the masses are lower when considering the effects of PV mixing as compared to the case when it is not considered. For the neutral $D$ mesons, the effects of PV mixing is significant. This is primarily due to the higher value of the $g_{PV}$ mixing parameter for the $D^0$ mesons. When PV mixing is considered, there is a massive drop in the mass of the pseudoscalar neutral meson as compared to when PV mixing is not considered. Similarly in \ref{fig1}c and \ref{fig1}d, there is increase in the mass of the vector $D$ mesons due to PV mixing. The masses of $D^{\pm}$ mesons increase with magnetic field due to the point particle correction with LLL contribution in the ground state energy. Around $4 m_\pi^2$, the effects of PV mixing becomes prominent for the charged mesons, unlike the neutral $D$ mesons, where the PV mixing effect is significant at non-zero magnetic field. For $D^{\pm}$ mesons, the contribution of PV mixing prevents the monotonous increase in mass with magnetic field at around $10 m_{\pi}^2$ where the mass plateaus.
\begin{figure}
    \centering
    \includegraphics[width=15cm]{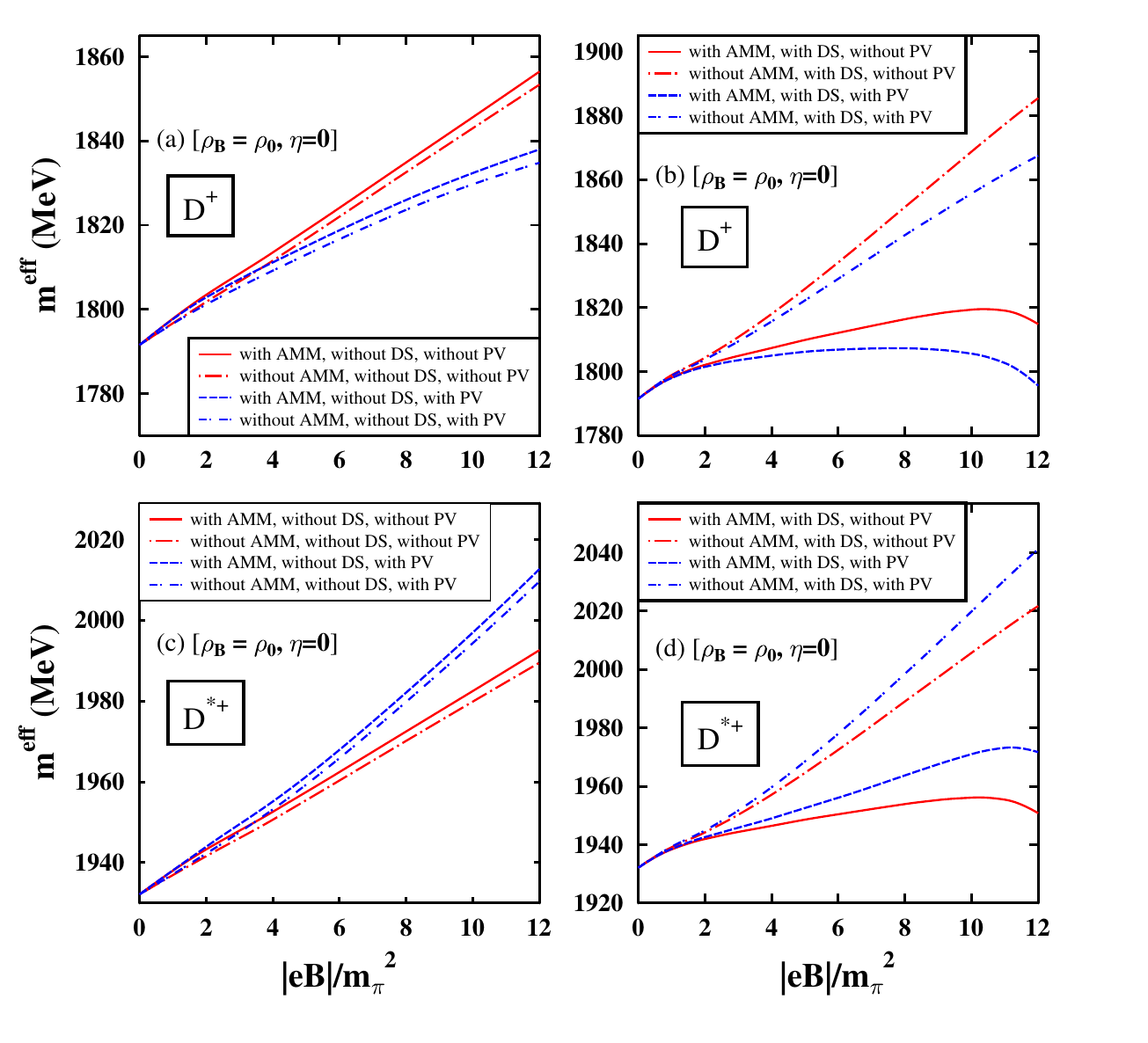}
\vspace{-0.8cm}
     \caption{The masses (in MeV) of $D^+$ and ${{D^*}^+}$ mesons are plotted as functions of $|eB|/m_\pi^2$ at $\rho_B=\rho_0$ for $\eta=0$, including the Dirac sea (DS) effects ((b) and (d)), which are compared to the case of without sea effects (in (a) and (c)). The masses are plotted with and without PV ($D^+-{{D^*}^{+||}})$ mixing effect, including the point-particle correction to the masses.} \label{fig2}
\end{figure}

In figure \ref{fig2}, the masses of the pseudoscalar $D^{+}$ mesons and the vector $D^{*+}$ mesons are plotted with increasing magnetic field at nuclear matter saturation density $\rho_B=\rho_0$ for symmetric matter ($\eta=0$). The masses of these mesons are a consequence of the combined effects of the Dirac sea contribution, nuclear matter contribution in external magnetic field and the PV mixing effects. In \ref{fig2}a and \ref{fig2}c, the DS effects have not been considered. The $D^+$ and $D^{*+}$ meson masses increase with increasing magnetic field due to the point particle correction. Due to $(D^+-D^{*+||})$ mixing, the masses of pseudoscalar meson $D^+$ are lower as compared to the case when PV mixing is not considered. Similarly the masses for the vector $D^{*+}$ mesons are larger when PV mixing is considered as compared to when it is not considered. In \ref{fig2}b and \ref{fig2}d, the DS effects have been considered. There are significant changes in mass spectrum when the AMM of the nucleons are considered. Without considering the AMM of the nucleons, the masses of the mesons increase with increasing magnetic field due to magnetic catalysis (MC) along with the LLL contribution to the masses. However for non-zero nucleonic AMM at finite density matter, the mass of the mesons gradually increase (instead of the steep increase as mentioned earlier) till $10 m_\pi^2$ and then decreases, as there is a competing effect due to inverse magnetic catalysis (IMC) against the point particle correction in this case. In both the above mentioned cases, PV mixing results in an overall decrease in mass for the pseudoscalar $D^+$ meson, and increase for the vector $D^{*+}$ meson. 

In figure \ref{fig4} the masses for the pseudoscalar $D^{-}$ mesons and the vector $D^{*-}$ mesons are plotted with increasing magnetic field at $\rho_B=\rho_0$ in the symmetric case. The trend in the variation of mass for the meson is similar to the previous case except for the difference in the absolute mass of the mesons. The mass of the $D^-$ and $D^{*-}$ is greater in nuclear matter than its antiparticles.

\begin{figure}
    \centering
    \includegraphics[width=15cm]{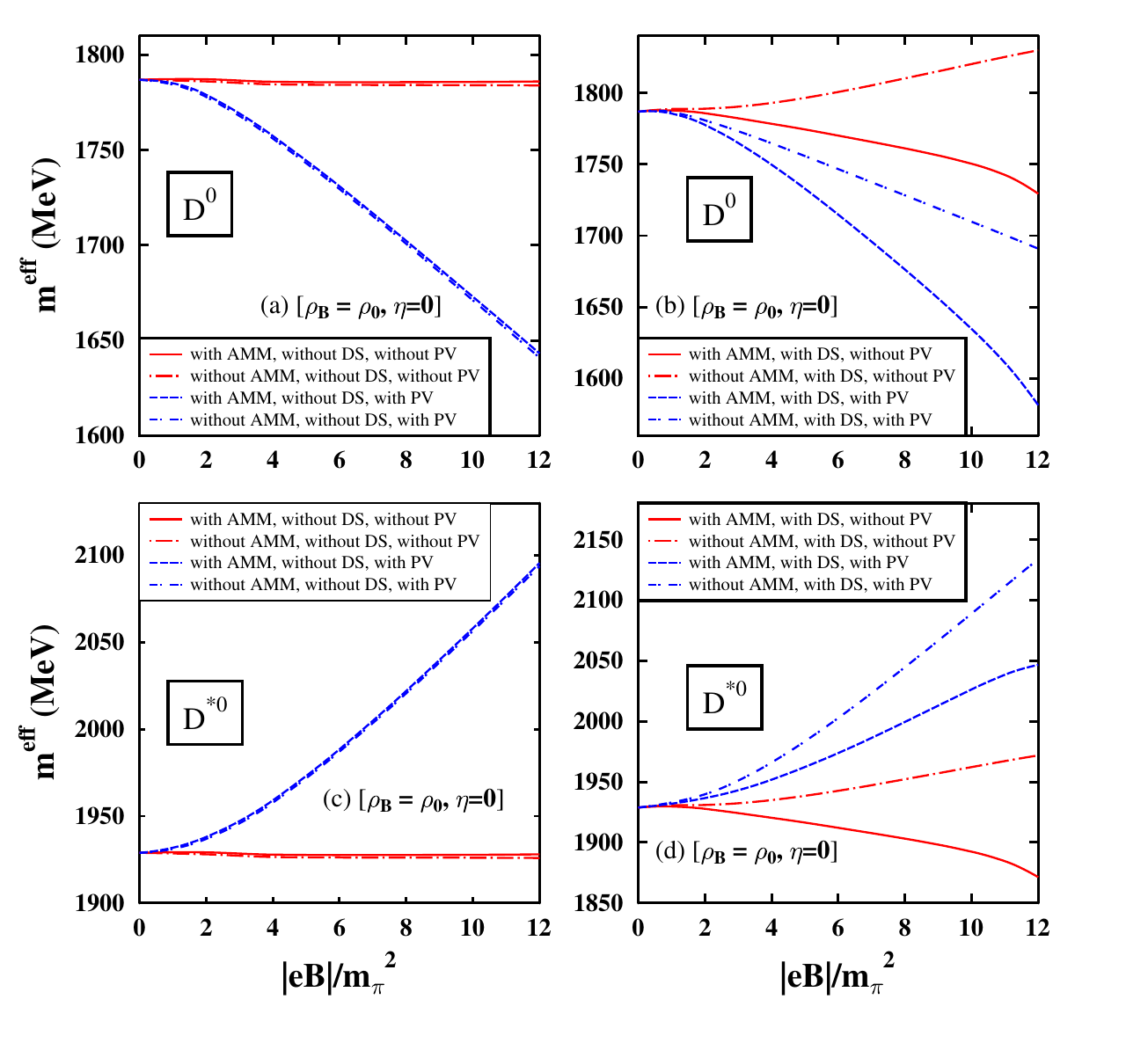}
\vspace{-0.8cm}
     \caption{The masses (in MeV) of $D^0$ and ${{D^*}^0}$ mesons are plotted as functions of $|eB|/m_\pi^2$ at $\rho_B=\rho_0$, ($\eta=0$), including the Dirac sea (DS) effects (in (b) and (d)). It is compared to the case when the DS effect is absent (in (a) and (c)). The masses are plotted with and without PV mixing of ($D^0-{{D}^{*0||}})$. } 
\label{fig3}
\end{figure}
\begin{figure}
    \centering
    \includegraphics[width=15cm]{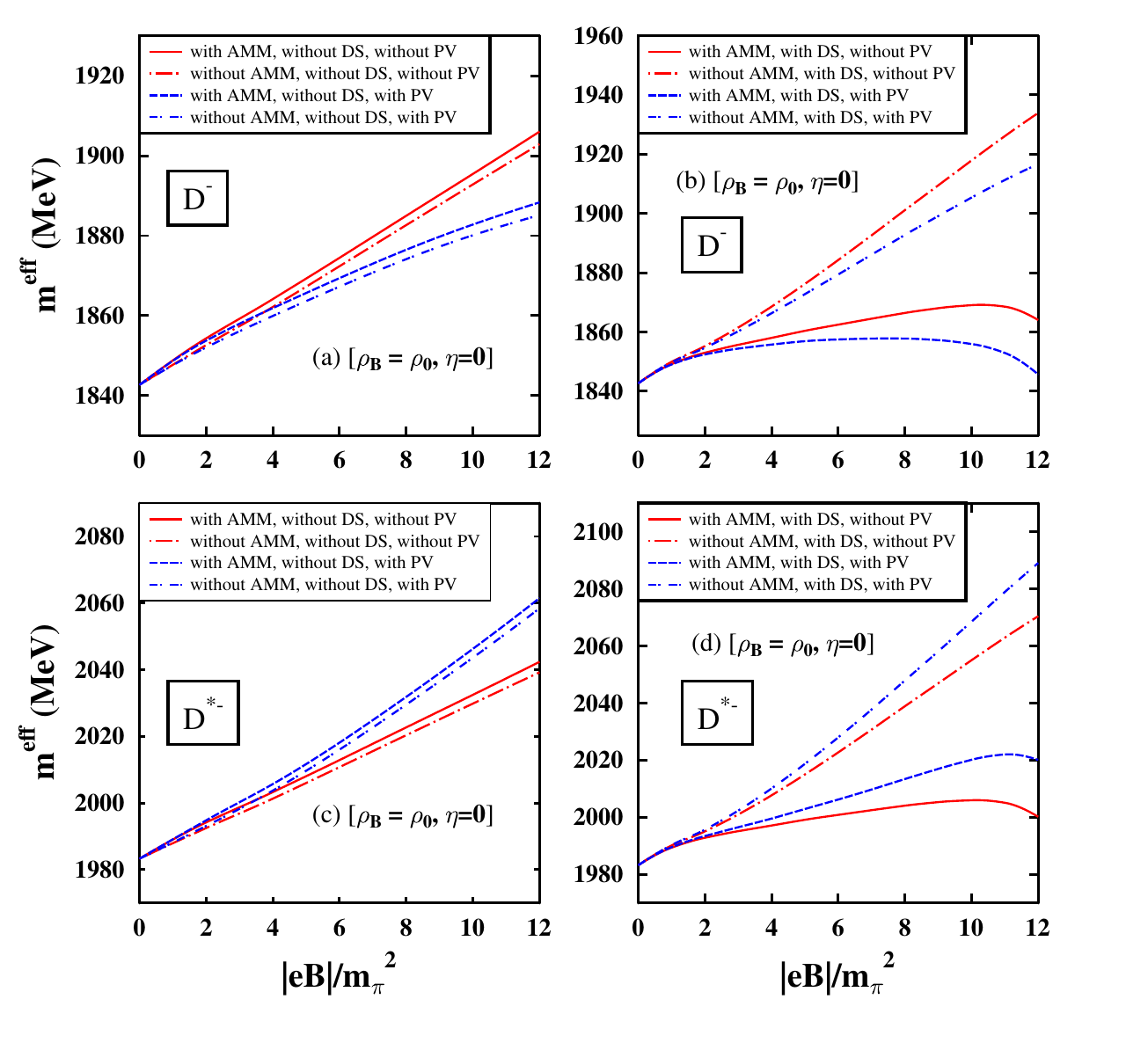}
\vspace{-0.8cm}
     \caption{The masses (in MeV) of $D^-$ and ${{D^*}^-}$ mesons are plotted as functions of $|eB|/m_\pi^2$ at $\rho_B=\rho_0$ ($\eta=0$),
including the Dirac sea (DS) effects (in (b) and (d)), are compared to the case with no sea effects (plots (a) and (c)). The masses are plotted with and without PV ($D^{-}-{{D^*}^{-||}})$ mixing effect, including the point-particle correction with lowest Landau level contribution.} 
\label{fig4}
\end{figure}
\begin{figure}[h!]
    \centering
    \includegraphics[width=15cm]{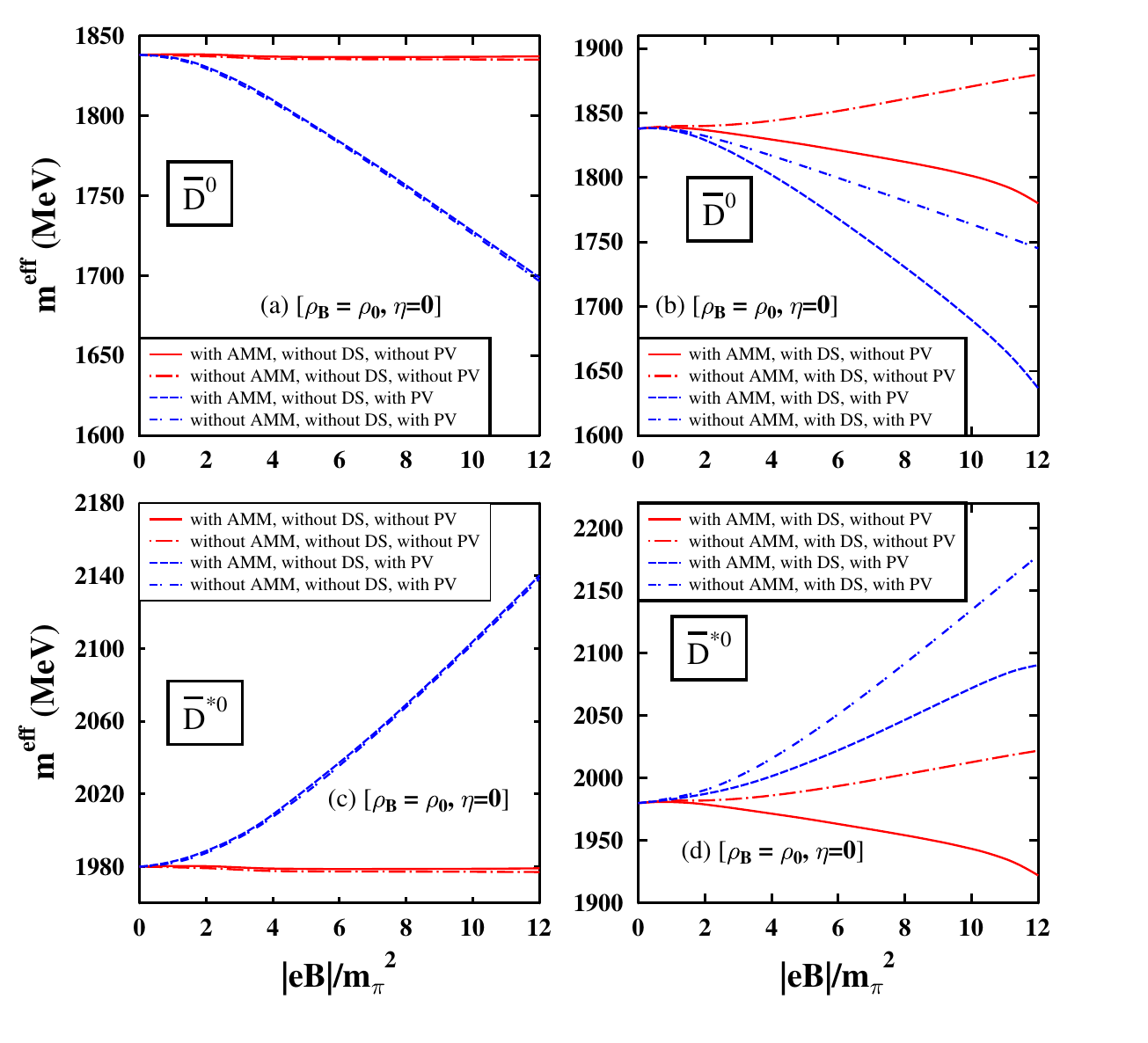}
\vspace{-0.8cm}
     \caption{The masses (in MeV) of $\bar{D}^0$ and $\bar{D}^{*0}$ mesons
are plotted at $\rho_B=\rho_0$ ($\eta=0$) as functions of $|eB|/m_\pi^2$, including the
Dirac sea (DS) effects (plots (b) and (d)). It is compared to
the case when the DS effect is absent (in (a) and (c)). The masses are plotted 
with and without PV mixing of ($\bar{D}^0-\bar{D}^{*0||})$.} 
    \label{fig5}
\end{figure}
Figure \ref{fig3} shows the variation of the mass for the neutral open charm mesons $(D^0,\  D^{*0})$ with magnetic field at $\rho_B=\rho_0$ and $\eta=0$. PV mixing results in large mass modifications for the neutral $D$ mesons due to the large value of the PV mixing parameter $g_{PV}$. The figures \ref{fig3}a and \ref{fig3}c, show the mass of the $D^0$ and $D^{*0}$ when PV mixing has been taken into account which is similar to the vacuum case except for the absolute masses due to their evaluation at nuclear matter. Figures \ref{fig3}b and \ref{fig3}d show the mass modifications for $D^0$ and $D^{*0}$ mesons with magnetic field accounting for the DS contributions. The mass modification for the mesons are significant when considering AMM of the nucleons as compared to when they are not considered. Besides, due to the lack of any contribution of the Landau energy level (because of charge neutrality), the effects of AMM of the nucleons is also significant when computing the neutral $D$ meson masses in nuclear matter. Considering the AMM of the nucleons, there is a gradual decrease in the masses of $D^0$ meson with magnetic field till $10 m_{\pi}^2$ after which the drop in mass is high, showing explicitly the effects of IMC on the masses when PV mixing is not considered. When AMM of the nucleons is not considered, the masses of the mesons increase with magnetic field due to MC at zero AMM in finite density matter. This occurs for both the $D^0$ and $D^{*0}$ mesons. When PV mixing is considered, there is an overall decrease in the mass of the pseudoscalar $D^0$ meson and an increase in the mass of the vector $D^{*0}$ meson.  In figure \ref{fig5} the masses for the pseudoscalar $\bar{D}^0$ mesons and the vector $\bar{D}^{*0}$ mesons are plotted with magnetic field at $\rho_B=\rho_0$ ($\eta=0$). The mass of the $\bar{D}$ mesons is greater than their antiparticles ($D$). The variation of masses of the above mentioned particles with magnetic field in isospin asymmetric nuclear matter with asymmetry parameter for e.g., $\eta=0.5$, are of similar pattern to the symmetric matter behavior, with slight changes in the values, hence the corresponding plots at $\rho_0$ and $\eta=0.5$, are not explicitly shown here. 

\begin{figure}
    \centering
    \includegraphics[width=15cm]{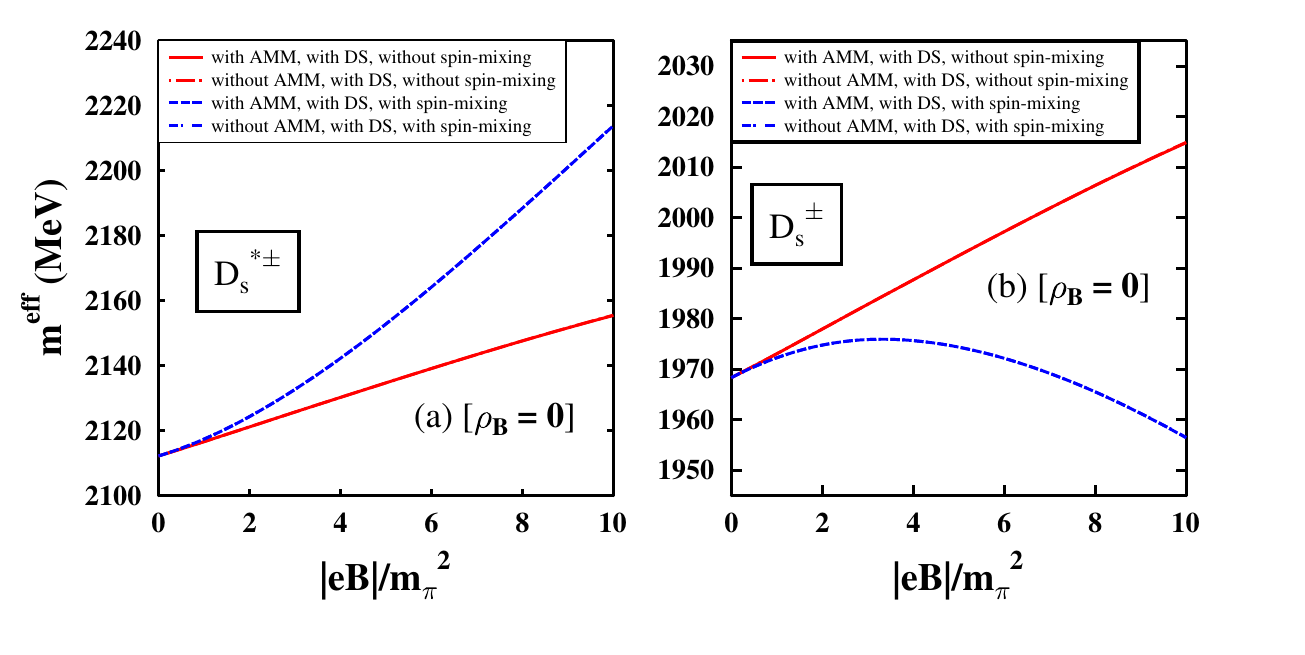}
\vspace{-0.8cm}
\caption{The masses (in MeV) of the $D_s^{\pm}$ and $D_s^{*\pm}$ mesons are plotted as functions of $|eB|/m_\pi^2$, at $\rho_B=0$, accounting for the Dirac sea (DS) effects and considering their spin-mixing ($D_s-{{D_s^*}^{||}})$ including the point-particle correction due to lowest Landau level.}
\label{fig6}
\end{figure}

\begin{figure}
    \centering
    \includegraphics[width=15cm]{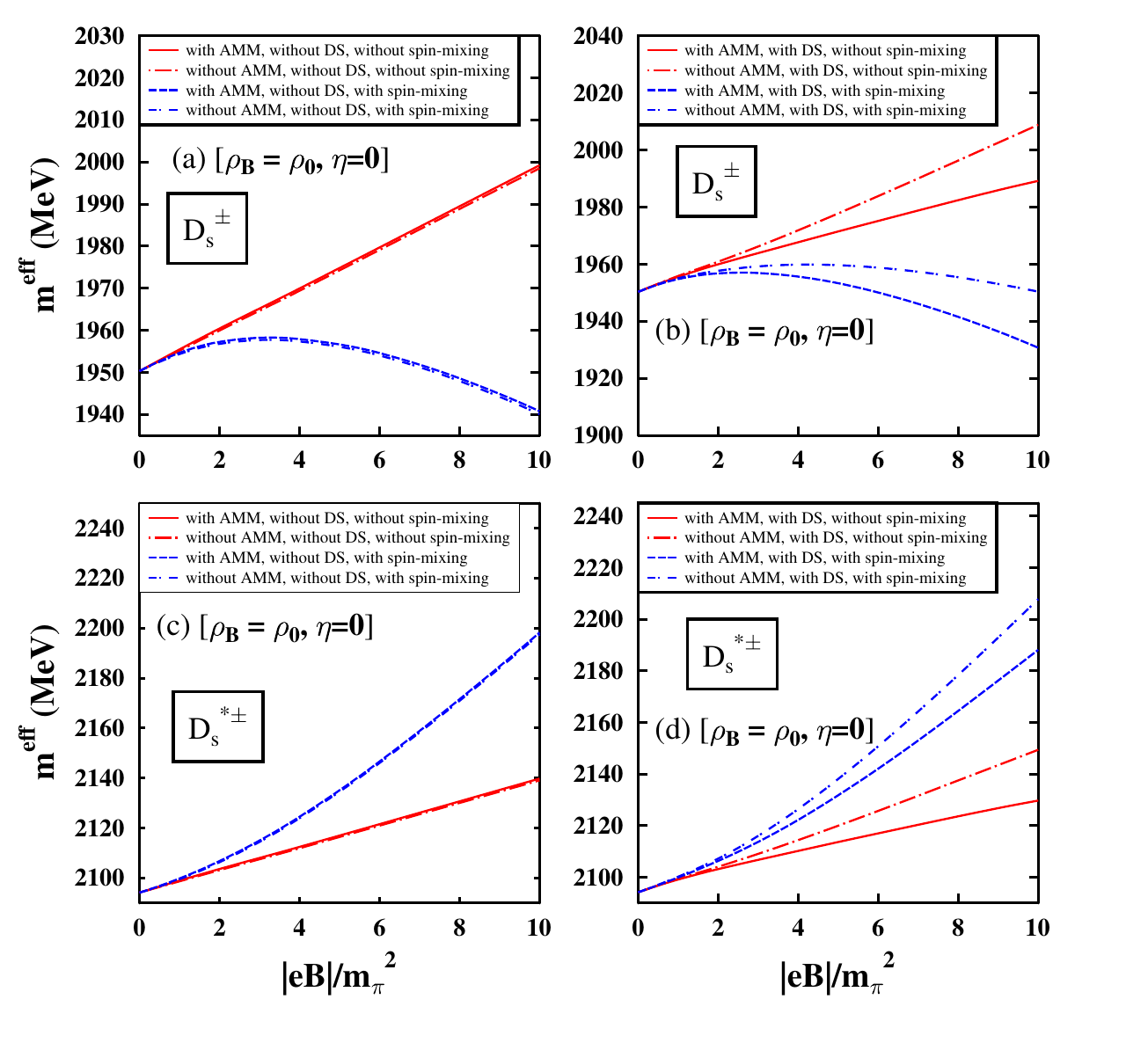}
\vspace{-0.8cm}
     \caption{The masses (in MeV) of $D_s^{\pm}$ and ${{D_s^*}^{\pm}}$ mesons
are plotted at $\rho_B=\rho_0$, $\eta$=0 as functions of $|eB|/m_\pi^2$, including the
Dirac sea (DS) effects (plots (b) and (d)), are compared to the masses with no DS effect (in (a) and (c)). The masses are plotted 
with and without the spin-mixing ($D_s-{{D_s^*}^{||}})$ effect, including the point-particle correction for the charged mesons. }
    \label{fig7}
\end{figure}

In figure (\ref{fig6}), the masses of the charmed, strange mesons are plotted as a function of magnetic field (in units of $m_{\pi}^2$), in the vacuum for $D_s^{*\pm}$ [in plot (a)] and $D_s$ [in plot (b)], accounting for the magnetized Dirac sea contribution. The effects of spin-magnetic field interaction between $D_s-D_s^{*\pm ||}$ lead to increasing (decreasing) mass of $D_s^{*\pm ||}$ ($D_s$) meson with magnetic field. In the absence of spin-mixing, the masses of the pseudoscalar and vector mesons increase with increasing magnetic field due to the effect of magnetic catalysis observed in vacuum for the case of without(w/o) and with AMM of the charged fermions in the Dirac sea contribution, and the additional point-particle correction for the ground state masses of these charged mesons in an external magnetic field.
The masses are plotted for finite baryon density at $\rho_B=\rho_0$ and $\eta=0$ in figure (\ref{fig7}). The comparison of Dirac sea contribution in the magnetic nuclear matter is shown to the case with no sea effect on the masses of the charmed strange mesons. The masses in plots (b) and (d), are seen to have a very slight increase for the case of non-zero AMM, w/o spin-mixing and with DS calculation, as compared to the masses in plots (a) and (c) with similar conditions. This is due to the fact that, in the calculation of Dirac sea contribution for magnetized matter with non-zero AMM, the values of the scalar fields decrease with magnetic field hence, one obtains the inverse magnetic catalysis effect, which along with the competing effect by Landau level contribution, lead to this slight increase in the masses. For the case of zero AMM, one obtains magnetic catalysis in magnetized nuclear matter, hence there obtained a sharp rise as compared to finite AMM case. The behavior of the masses of $D_s$ and $D_s^{*}$ mesons due to spin-mixing are similar as that described in vacuum. Therefore, the observed behavior is a combined effect of (inverse) magnetic catalysis, point-particle correction and level repulsion due to spin-mixing of $D_s-D_s^{*||}$ mesons. Similar variation with slight change in the values are obtained for the asymmetric matter case.    
\begin{figure}
    \centering
    \includegraphics[width=15cm]{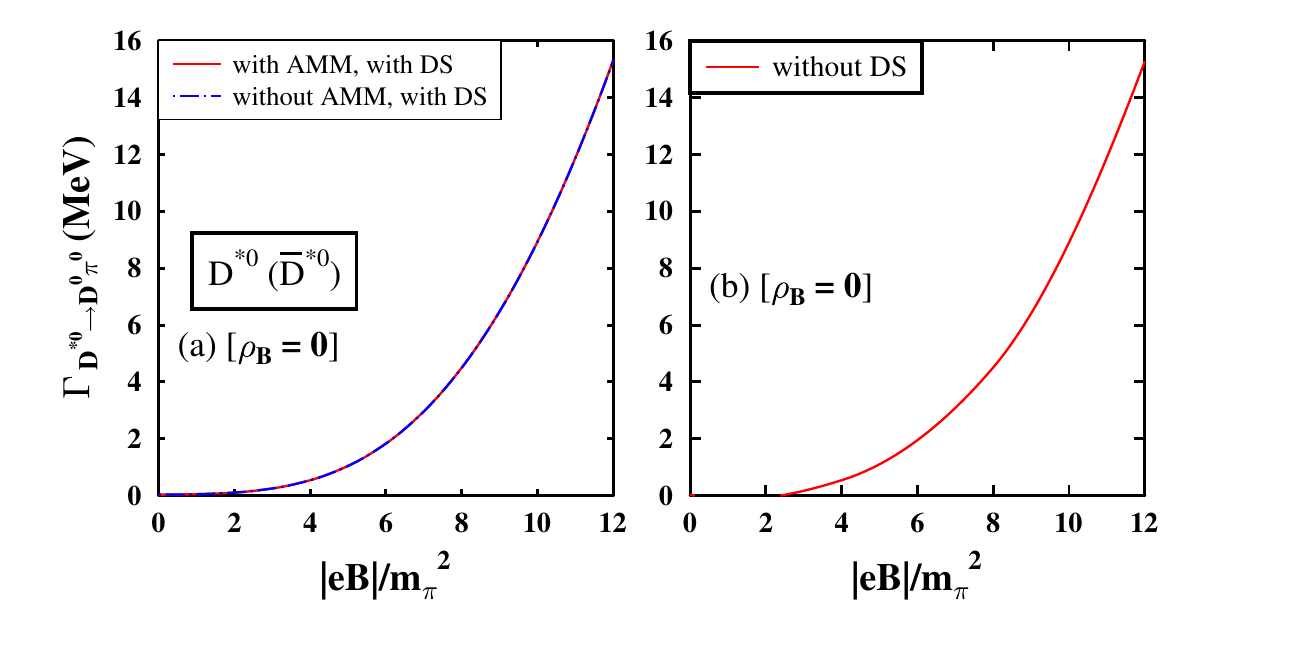}
\vspace{-0.8cm}
     \caption{In-medium partial decay widths (in MeV) of $D^{*0}\rightarrow D^0\pi^0$ [$\bar{D}^{*0} \rightarrow \bar{D}^0\pi^0 $] are plotted as functions of $|eB|/m_{\pi}^2$, at $\rho_B = 0$, considering the Dirac sea (DS) effects. The decays incorporate the PV mixing of $(D^0 - D^{*0||})$ and $(\bar{D}^0-\bar{D}^{*0||})$, along with the non-PV mixed, transverse components mass. The DS effect in (a), is compared to case when there is no DS effect [(b)].}
    \label{fig8}
\end{figure}
\begin{figure}[h!]
    \centering
    \includegraphics[width=15cm]{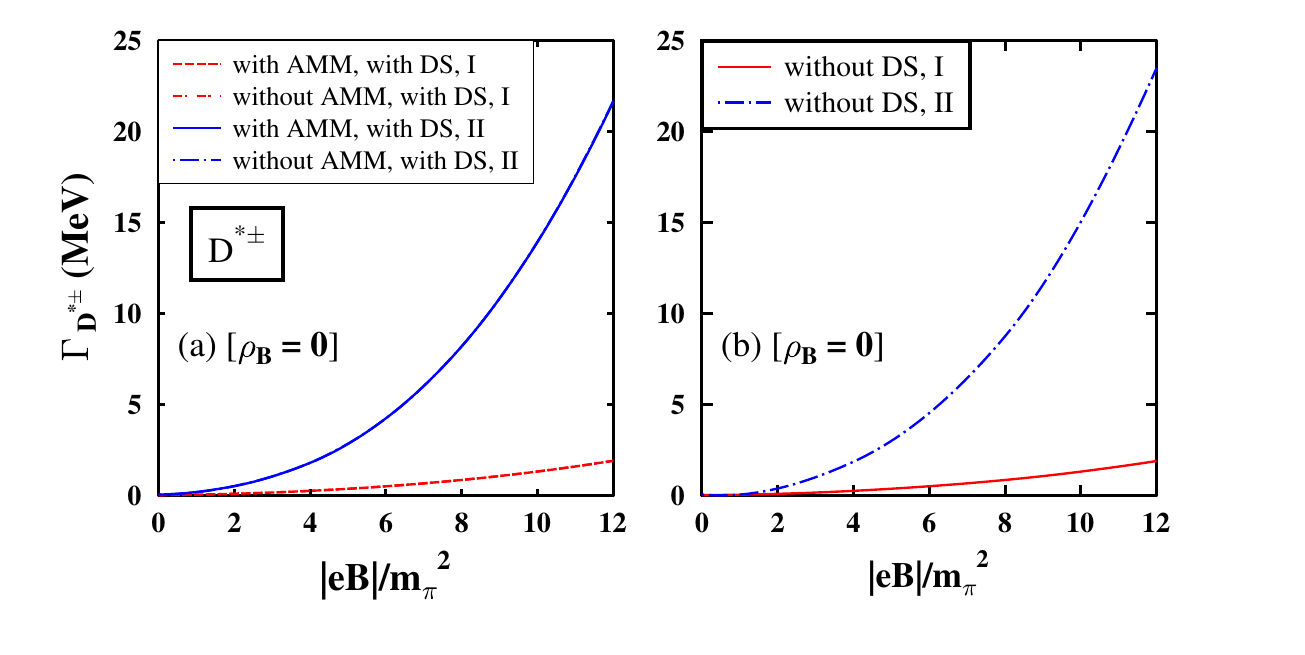}
\vspace{-0.8cm}
     \caption{In-medium partial decay widths (in MeV) of $D^{*\pm}\rightarrow D^{\pm}\pi^0$ (I), $D^{*\pm}\rightarrow D^0\pi^{\pm}$ (II) are plotted as functions of $|eB|/m_{\pi}^2$, at $\rho_B = 0$, accounting for the Dirac sea (DS) effects. The decays incorporate the PV mixing of $(D^{\pm} - D^{*\pm||})$ and $(D^0-D^{*0||})$, along with the non-PV mixed transverse components masses for the charged vector mesons. }
    \label{fig9}
\end{figure}
\begin{figure}
    \centering
    \includegraphics[width=15cm]{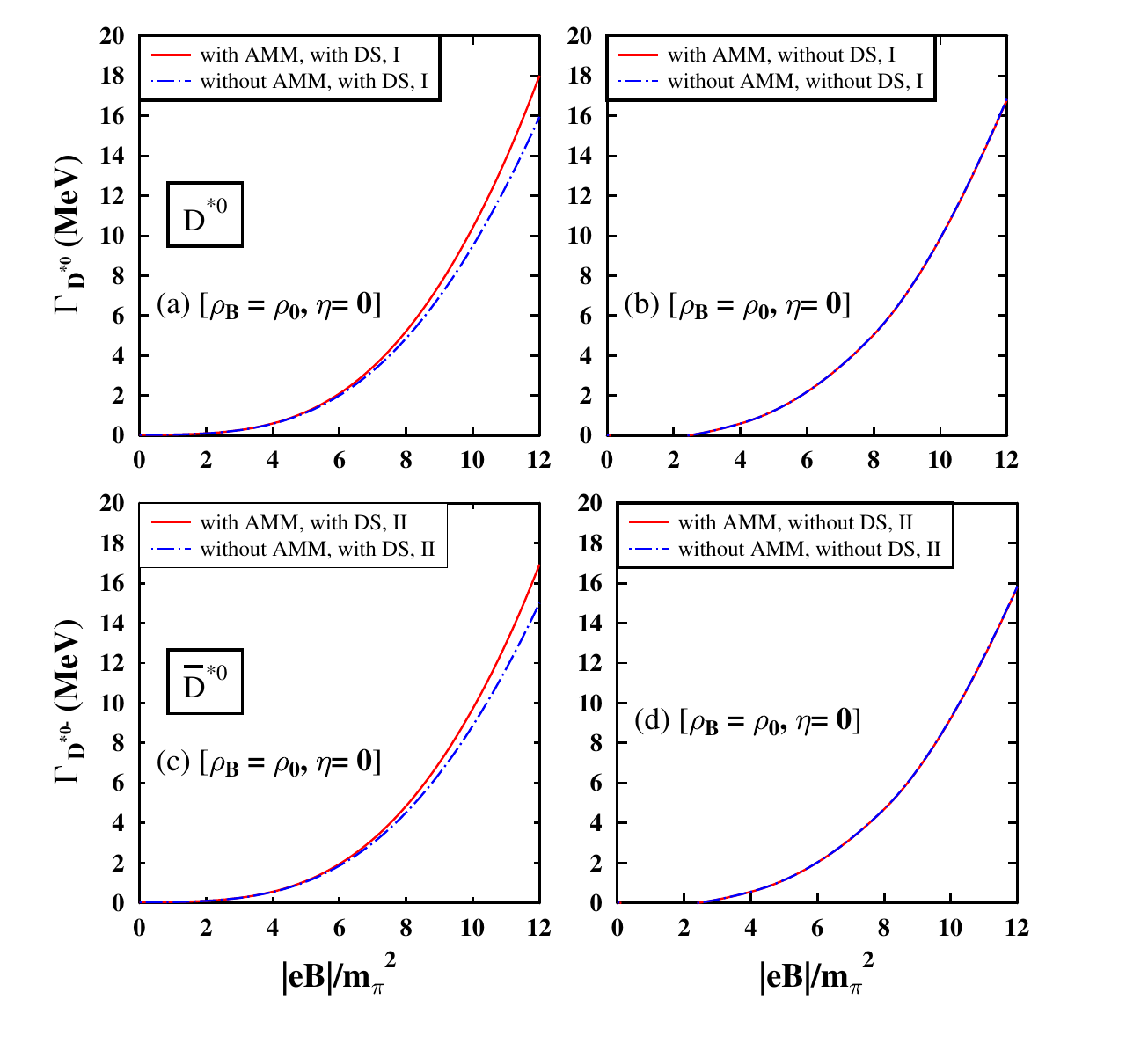}
\vspace{-0.8cm}
     \caption{The decay widths (in MeV) of $D^{*0}\rightarrow D^0\pi^0$ (I) and $\bar{D}^{*0} \rightarrow \bar{D}^0\pi^0 $ (II) are plotted as functions of $|eB|/m_{\pi}^2$, at $\rho_B = \rho_0$ and $\eta=0$, accounting for the Dirac sea (DS) effects. The decays incorporate the PV mixed masses due to mixing of $(D^0 - D^{*0||})$ and $(\bar{D}^0-\bar{D}^{*0||})$. }
    \label{fig10}
\end{figure}

Figure \ref{fig8} shows the in-medium partial decay widths of $D^{*0} \rightarrow D^0 \pi^0 (\bar{D}^{*0} \rightarrow \bar{D} \pi^0)$ with magnetic field at zero density incorporating the DS effects. The decays incorporate the PV mixing effects of $(D^0 - D^{*0||})$ and $(\bar{D}^0-\bar{D}^{*0||})$, along with the non-PV mixed, transverse components mass. The DS effects are considered in \ref{fig8}a as compared to the no DS effects in \ref{fig8}b. The effect of AMM of the charged fermions is not significant at zero density through DS calculation. The decay widths are $0.5,\ 1.80,\ 4.25,\ 8.21,\ 13.94,\ 21.65$ (in MeV) at $|eB|/m_{\pi}^2=2,\ 4,\ 6,\ 8,\ 10,\ 12$. In figure \ref{fig9}, the in-medium partial decay widths (in MeV) of $D^{*\pm}\rightarrow D^{\pm}\pi^0$ (I), $D^{*\pm}\rightarrow D^0\pi^{\pm}$ (II) are plotted as functions of $|eB|/m_{\pi}^2$, at $\rho_B = 0$, accounting for the Dirac sea (DS) effects. The decays incorporate the PV mixing effects of $(D^{\pm} - D^{*\pm||})$ and $(D^0-D^{*0||})$, along with the non-PV mixed transverse components masses for the charged vector mesons. The Dirac sea effect in (a) is compared to the no DS effect in (b), with and without the effects of the AMMs. The decay widths for channel I are $0.025,\ 0.09,\ 0.26,\ 0.51,\ 0.86,\ 1.32,\ 1.91$ MeV at $0,\ 2,\ 4,\ 6,\ 8,\ 10,\ 12$ $ |eB|/m_{\pi}^2$ for non-zero AMM. For channel II, the decay widths are $0.056,\ 0.52,\ 1.80,\ 4.25,\ 8.21,\ 13.94,\ 21.65$ MeV at $0,\ 2,\ 4,\ 6,\ 8,\ 10,\ 12$ $ |eB|/m_{\pi}^2$ at finite AMM. The values of decay widths without AMM are similar to what is obtained when AMM is included. Clearly channel II has a greater decay width than channel I. This is due to the fact that channel I has $D^+$ as its daughter particle which gets mass contribution from the LLL and PV mixing. Channel II has $D^0$ as daughter particle and receives negative mass shift due to the PV mixing. Hence channel II has a greater decay width than channel I.  

Figure \ref{fig10} shows the decay widths (in MeV) of $D^{*0}\rightarrow D^0\pi^0$ (I) and $\bar{D}^{*0} \rightarrow \bar{D}^0\pi^0 $ (II) are plotted as functions of $|eB|/m_{\pi}^2$, at $\rho_B = \rho_0$ and $\eta=0$, accounting for the Dirac sea effects. The decays incorporate the PV mixed masses due to mixing between $(D^0 - D^{*0||})$ and $(\bar{D}^0-\bar{D}^{*0||})$. The Dirac sea effects (with DS), in (a) and (c) are compared to the no DS effect in (b) and (d), with and without the effects of the nucleons AMMs. The decay widths increase with magnetic field. The decay widths for (I) with (without) AMM are $0.035\ (0.035),\ 0.61\ (0.59),\ 5.23\ (4.87),\ 18.0\ (15.94)$ MeV at $0,\ 4,\ 8,\ 12$ $ |eB|/m_{\pi}^2$. The decay widths for (II) with (without) AMM are $0.036\ (0.036),\ 0.56\ (0.56),\ 4.86\ (4.53),\ 16.94\ (15.0)$ MeV at similar values of magnetic field.

\begin{figure}
    \centering
    \includegraphics[width=15cm]{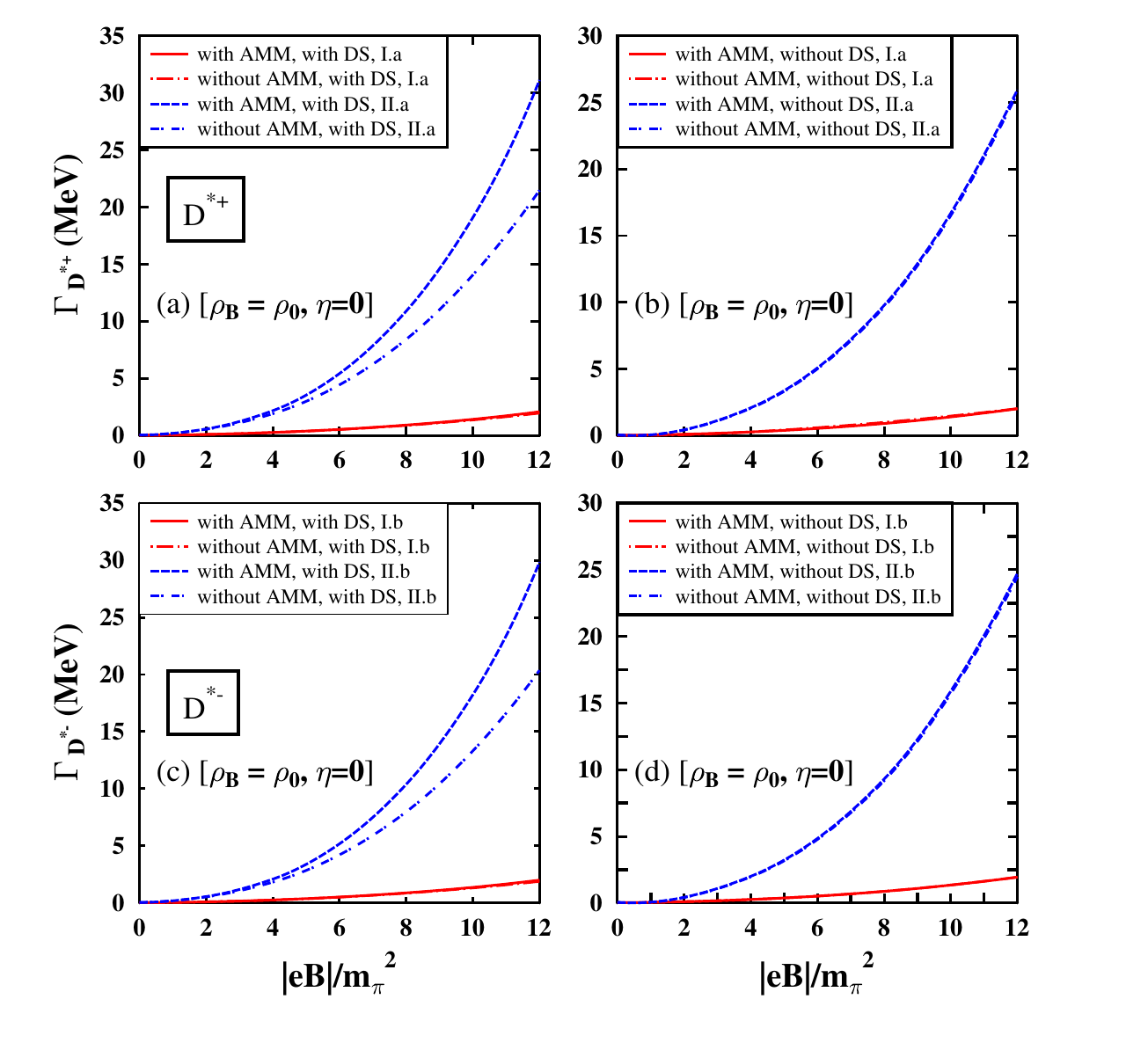}
\vspace{-0.8cm}
     \caption{In-medium partial decay widths (in MeV) of $D^{*+}\rightarrow D^+\pi^0$ (I.a), $D^{*+}\rightarrow D^0\pi^+$ (II.a) and $D^{*-}\rightarrow D^-\pi^0$ (I.b), $D^{*-}\rightarrow \bar{D}^0\pi^-$ (II.b) are plotted as functions of $|eB|/m_{\pi}^2$, at $\rho_B = \rho_0$ ($\eta=0$), accounting for Dirac sea (DS) effects. Decays incorporate the PV mixing effects of $(D^{\pm} - D^{*\pm||})$. }
    \label{fig11}
\end{figure}

Figure \ref{fig11} shows in-medium partial decay widths (in MeV) of $D^{*+}\rightarrow D^+\pi^0$ (I.a), $D^{*+}\rightarrow D^0\pi^+$ (II.a) and $D^{*-}\rightarrow D^-\pi^0$ (I.b), $D^{*-}\rightarrow \bar{D}^0\pi^-$ (II.b) are plotted as functions of $|eB|/m_{\pi}^2$, at $\rho_B = \rho_0$ and $\eta=0$, accounting for the Dirac sea effects. The decays incorporate the PV mixing effects of $(D^+ - D^{*+||})$ and $(D^- - D^{*-||})$. The Dirac sea effects (with DS), in (a) and (c) are compared to the no DS effect in (b) and (d), with and without the effects of the nucleons' AMM. The decay widths increase with magnetic field. The decay widths for (Ia) with (without) AMM are $0.025\ (0.025),\ 0.27\ (0.27),\ 0.92\ (0.89),\ 2.09\ (1.97)$ MeV at $0,\ 4,\ 8,\ 12$ $ |eB|/m_{\pi}^2$. For channel (Ib), the decay widths with (without) AMM are $0.025\ (0.025),\ 0.26\ (0.26),\ 0.88\ (0.86),\ 2.0\ (1.89)$ MeV at $0,\ 4,\ 8,\ 12$ $ |eB|/m_{\pi}^2$. Channels (IIa) and (IIb) have higher decay widths compared to (Ia) and (Ib), due to the large reduction in mass experience by the $D^0(\bar{D}^0)$ mesons. The decay widths for (IIa) are $0.05\ (0.05),\ 2.19\ (1.92),\ 10.89\ (8.42),\ 31.12\ (21.47)$ MeV with (without) AMM of nucleons at $0,\ 4,\ 8,\ 12$ $ |eB|/m_{\pi}^2$. Similarly for (IIb), the decay widths are $0.05 \ (0.05),\ 2.09\ (1.83),\ 10.39\ (7.98),\ 29.84\ (20.37)$ MeV at $0,\ 4,\ 8,\ 12$ $ |eB|/m_{\pi}^2$ with (without) AMM.

\begin{figure}
    \centering
    \includegraphics[width=15cm]{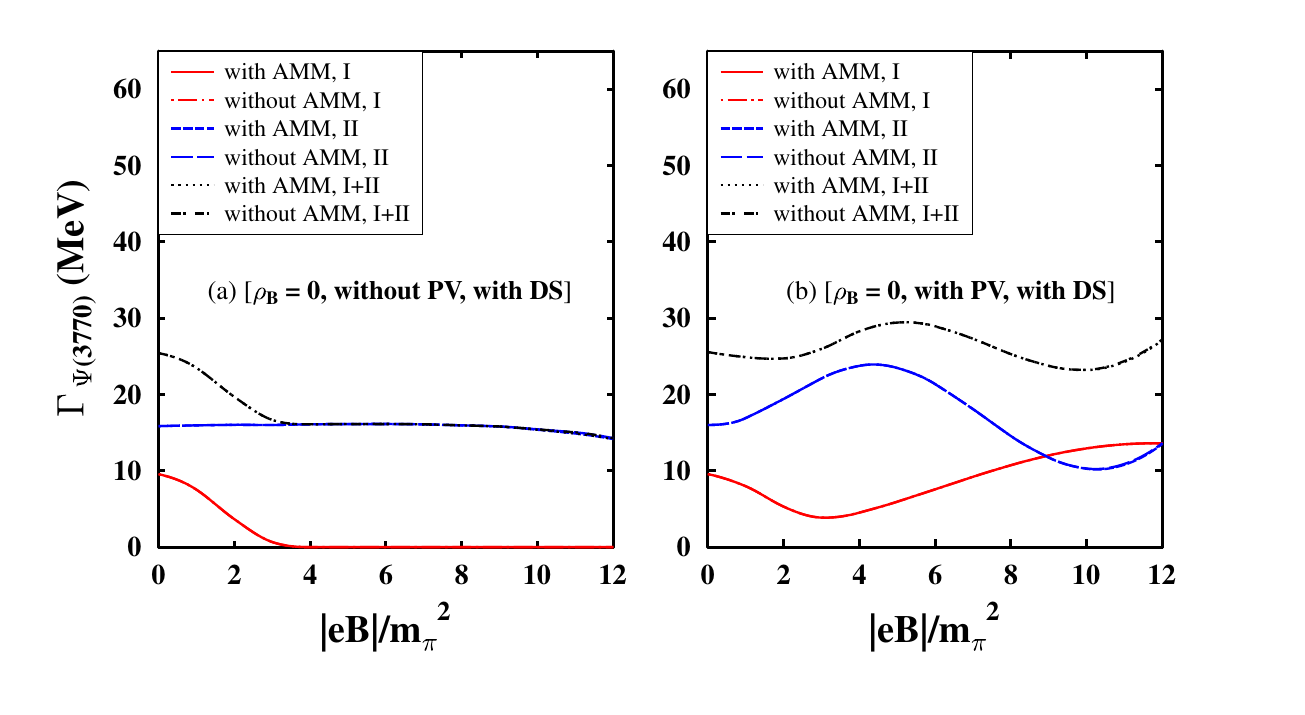}
\vspace{-0.8cm}
     \caption{The in-medium partial decay widths (in MeV) of $\Psi(3770)\rightarrow D^+D^-$ (I) and $\Psi(3770)\rightarrow D^0\bar{D}^0$ (II) and the total (I+II), are plotted as functions of $|eB|/m_{\pi}^2$, at $\rho_B = 0$, including the Dirac sea (DS) effects. The PV mixing effects on the parent and daughter particle states are also considered. }
    \label{fig14}
\end{figure}
\begin{figure}[h!]
    \centering
    \includegraphics[width=15cm]{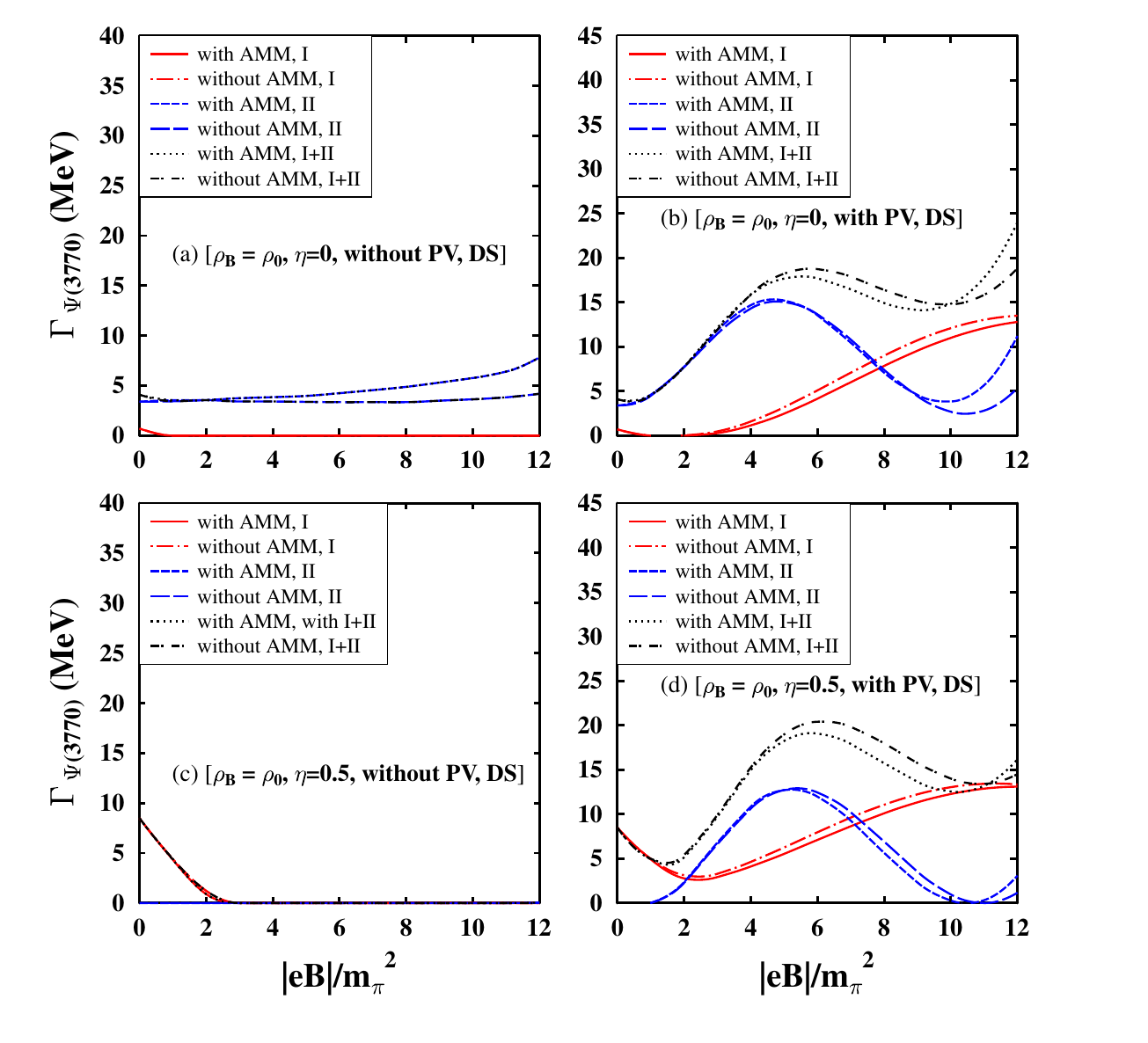}
\vspace{-0.8cm}
     \caption{The decay widths (in MeV) of $\Psi(3770)\rightarrow D^+D^-$ (I) and $\Psi(3770)\rightarrow D^0\bar{D}^0$ (II) and the total (I+II), are plotted as functions of $|eB|/m_{\pi}^2$, at $\rho_B = \rho_0$ and $\eta=0,\ 0.5$, including the Dirac sea (DS) effects. The PV mixing effects on the parent and daughter particle states are also considered.}
    \label{fig15}
\end{figure} 
 
In figure \ref{fig14}, the decay widths of $\Psi(3770) \rightarrow D^+ D^-$ (I) and $\Psi(3770) \rightarrow D^0 \bar{D}^0$ (II) are plotted with variation in the magnetic field at $\rho_B=0$, accounting for the Dirac sea effects. In plot (a), PV mixing effect on the masses of the parent and daughter particles is not considered and in plot (b) it is considered.
%In case of plot (b), the contributions from the PV mixed $S_z=0$ component and non-PV mixed transverse components for vector charmonium state have been considered separately.
For channel I, the daughter particles are the charged $D^{\pm}$ meson. These particles receive additional effect due to their point particle correction to the ground state energy in an external magnetic field. Besides, their vacuum mass is greater than the daughter particles $D^0 \bar{D}^0$ corresponding to channel II. As such, the decay width for channel I is lower than channel II. At around $4 m_{\pi}^2$ the sum of the product particles' masses in channel I becomes larger than the parent meson. Hence, at higher  magnetic field, the decay width for this channel becomes zero. The total decay width (I+II) is plotted. Plot (b) shows the decay width for I and II when PV mixing is taken into account. For channel I, the decay width first decreases around $4 m_{\pi}^2$ and then increases with magnetic field. This is due to the fact that initially at low magnetic field, the LLL contribution dominates the mass of $D^+D^-$. At higher magnetic field, the PV mixing dominates and the decay width increases. For channel II, the decay width increases till $4 m_{\pi}^2$, then decreases. At around $10 m_\pi^2$ it increases again. 

%In figure \ref{fig15}, the decay widths are plotted at $\rho_B=\rho_0$, $\eta=0$, accounting for the Dirac sea in plots (a) and (c). This is compared to the case when DS effect on the masses is absent, in plots (b) and (d). The various effects considered are shown in the plots. 

%In this study the in-medium partial decay widths of the open charm and charmonium states have been calculated in the magnetized nuclear matter incorporating the effects of the magnetized Dirac sea along with the PV mixing on the parent and daughter particles' masses, as have been illustrated in this section. The significant effects of the (inverse) magnetic catalysis on the in-medium properties of the open charm and charmonium states thus obtained, should modify the production and propagation of the charmonia and open charm mesons in the peripheral ultra relativistic heavy ion collision experiments. 

In figure \ref{fig15} the decay widths of $\Psi(3770)\rightarrow D^+D^-(I)$ and $\Psi(3770)\rightarrow D^0\bar{D}^0(II)$ and the total decay width encompassing the two channels are plotted as function of $|eB|/m_{\pi}^2$ at $\rho_B=\rho_0$ at $\eta=0,0.5$ incorporating the Dirac sea effects. Figures \ref{fig15}a and \ref{fig15}c show the decay widths when PV mixing has not been considered, figures \ref{fig15}b and \ref{fig15}d show the effects of PV mixing. For isospin asymmetry parameter $\eta=0$, without PV, in fig.\ref{fig15}a, the channel $I$ decay width is 0.7 MeV at $|eB|/m_{\pi}^2=0$. With increasing magnetic field, it remains zero irrespective of whether AMM of the nucleons is considered or not. For channel $II$, the values of decay widths vary as $3.39,\ 3.84,\ 4.24,\ 5.76$ MeV at $0,\ 4,\ 8,\ 10$ $ |eB|/m_{\pi}^2$ when AMM is considered; $3.39,\ 3.41,\ 3.34,\ 3.63$ MeV at $0,\ 4,\ 8,\ 10$ $ |eB|/m_{\pi}^2$ when AMM is not considered. Thus when AMM is considered, the masses of the neutral $D$ undergo slight decrease in mass as compared to when AMM is not considered (discussed previously) and have slightly higher value of decay widths. The total decay width is obtained by adding the values of decay widths for individual channels. Fig.\ref{fig15}c shows that the decay widths for channel $I$ is greater than channel $II$ around $2 eB/m_{\pi}^2$. For channel $I$, the decay widths are $8.49,\ 4.33,\ 0.89$ MeV at  $0,\ 1,\ 2$ $ |eB|/m_{\pi}^2$ when AMM is considered; $8.49,\ 4.39,\ 1.22$ MeV at $0,\ 1,\ 2$ $ |eB|/m_{\pi}^2$ when AMM is not considered. For channel $II$, the decay width is zero irrespective of the changing magnetic field due to the variation of mass of neutral $D$ meson with magnetic field (discussed earlier). In figure\ref{fig15}b, the decay widths are plotted when the effects of PV mixing are considered, thereby showing significant variation in decay width with magnetic field. For channel I, the decay width varies as $1.15,\ 7.84,\ 12.79$ MeV at $4,\ 8,\ 12$ $ |eB|/m_{\pi}^2$ (with AMM) indicating increasing decay width. This occurs due to the $D^{\pm}$ meson and $\Psi(3770)$ mass modification (due to PV mixing with $\eta_c(2S)$). For channel II, the decay width varies as $7.75,\ 14.67,\ 13.54,\ 7.07,\ 3.84,\ 11.21$ MeV at $2,\ 4,\ 6,\ 8,\ 10,\ 12$ $ |eB|/m_{\pi}^2$ (with AMM). Without AMM, the decay width variation is similar till $9 m_{\pi^2}$ after which the variation is significant; $2.68,\ 5.34$ MeV at $10,\ 12$ $ |eB|/m_{\pi}^2$. Figure\ref{fig15}d shows the decay widths for purely neutron matter $(\eta=0.5)$ with PV mixing. The decay widths for channel I (with AMM) are $8.49,\ 2.77,\ 7.10,\ 12.26,\ 13.09$ MeV at $0,\ 2,\ 6,\ 10,\ 12$ $ |eB|/m_{\pi}^2$. For channel II the decay widths are $0,\ 2.36,\ 10.84,\ 11.96,\ 5.65,\ 0.28,\ 3.07$ MeV at 
 $0,\ 2,\ 4,\ 6,\ 8,\ 10,\ 12$ $ |eB|/m_{\pi}^2$, showing nodal behaviour (decay width approaches zero at certain magnetic fields). The values for decay widths without AMM, are similar to its (with AMM) counterpart. For figure \ref{fig15}b and \ref{fig15}d, the total decay widths are obtained by adding the values of two channels.

Heavy quark (charm and bottom), due to their large mass are produced in the initial stages of high energy heavy ion collisions (e.g., ALICE in LHC) via hard partonic scattering process and serve as very good probes of the produced medium. Precise measurements of yield modifications of the heavy flavor hadrons in heavy ion collisions along with other factors give rise to constraints on transport coefficients of the strongly coupled medium. However, in non-central ultra-relativistic heavy ion collision experiments at RHIC, LHC, a strong magnetic field have been estimated to be produced at the early stages of collisions \cite{kharzeev, skokov}, which eventually decrease with time depending on the electrical conductivity of the produced medium and needs solutions of the magnetohydrodynamic equations \cite{marasinghe}. The study of the in-medium hadronic properties under a static magnetic field can be executed if the distance over which the magnetic field varies considerably, is much bigger than the size of the hadrons. The effect of magnetic field on the dissociation rate of produced quarkonium in the central rapidity region of relativistic heavy ion collisions have been studied and it has been observed that the quarkonium dissociation energy increases with increasing magnetic field \cite{marasinghe} and with quarkonium momentum. In recent heavy flavor measurements from ALICE in pp, p-Pb, Pb-Pb collision systems, production of charmed mesons are described by QCD-inspired model within uncertainties, in which ratio of different particle species, for e.g., of $D_s/D_0$ is sensitive to the hadronisation mechanism in vacuum and in medium. Ratio of $D$ mesons is found to be universal among different colliding systems and center-of-mass energies unlike the baryon-to-meson ratios which are not illustrated by universal mechanism in different collider systems.
Mass modification (in \%) of the neutral $D^0$ mesons at zero density is about $0.01\ (-0.36), 0.04\ (-1.38), 0.07\ (-2.67), 0.08\ (-4.13)$ when PV mixing is considered (not considered) at $|eB|/m_{\pi}^2=2, 4, 6, 8$. The values for $D_s$ mesons are $0.48 (0.33),\ 0.98 (0.37),\ 1.47 (0.19),\ 1.93 (-0.14)$ percent when the LLL approximation is used with no PV mixing (PV mixing) included at $|eB|/m_{\pi}^2=2,\ 4,\ 6,\ 8$. In the present work, we have studied the Dirac sea contribution over a range of magnetic field from $|eB|=0$ to 12$m_{\pi}^2$ at $\rho_B=0$ and $\rho_0$, which should cover the dynamics at the freeze-out hypersurface (FOHS) where the estimated magnetic field is low, around $0.07$ Ge$V^2$ \cite{fohs}, hence may give rise to observable impacts as illustrated here. For the $D$ meson spectra in LHC, a larger suppression is observed at low to intermediate transverse momentum and is larger in the most centrality class, as compared to the semi-peripheral centrality class, due to the final state effects as measured by the nuclear modification factor in 
$p-Pb$ collisions. These results are compatible with model considering cold nuclear matter effects \cite{alice}, hence the effects of magnetized matter on the properties of heavy flavor hadrons specifically on the charm mesons are of considerable phenomenological importance. 
The medium modified masses of the open charm mesons $D^0$, $D_s$ as studies in our present work in the magnetic matter may thus have important observable impacts in the particle production ratio of $D_s/D_0$, from the recent measurements of relative abundances of such particle species in ALICE measurements \cite{alice}. The medium modifications of the excited state of charmonium $\Psi(3770)$ may also lead to in-medium partial decay widths to open charm mesons $D\bar{D}$ which further affect the yield of open charm and lowest lying states of charmonia, for e.g., $J/\psi$. The medium modifications of open charm and $\Psi(3770)$ could have observable impacts for e.g., on $J/\psi$ suppression in the relativistic heavy ion collision experiments.

\section{Summary}
\label{sec5}
To summarize, we have investigated the in-medium masses of the pseudoscalar open charm ($D$ and $\bar D$) and the charmed, strange ($D_s^{\pm}$) mesons, as well as the vector, open charm (${D^*}$ and $\bar {D}^*$) and the charmed, strange ($D_s^{*\pm}$) mesons, in the magnetized nuclear
matter accounting for the contributions of the Dirac sea. The effects of PV mixing between the pseudoscalar and longitudinal component of the vector mesons are considered in the present study. In an external magnetic field, the contribution of the lowest Landau energy level as a point-particle correction, is considered in the mass calculation of the charged open charm and charmed, strange  mesons. The Dirac sea (DS) contribution is observed to lead to an enhancement of the light quark condensates with $|eB|$, an effect called magnetic catalysis, at zero baryon density, for non zero and zero AMM of the charged fermions. The light quark condensates at the nuclear matter saturation density, $\rho_0$, for the non-zero AMMs of nucleons, decrease with increasing $|eB|$ due to the Dirac sea contributions, which lead to an inverse magnetic catalysis effect. However, for vanishing AMM of the nucleons, opposite behavior is observed, at $\rho_0$. The effects of the (inverse) magnetic catalysis on the mass of the open charm mesons are observed to be significant in comparison to the case when DS contribution is not taken into account. There are observed to be appreciable effects of the anomalous magnetic moments through the magnetized Dirac sea at finite density matter, whereas there is almost no difference found at $\rho_B=0$ based on the AMM of charged fermions in DS calculation. In the presence of an external magnetic field, important effects are obtained due to the PV mixing, which leads to the drop (rise) in the masses of the pseudoscalar (longitudinal component of the vector) mesons, i.e., of $D(D^{*||})$, $\bar{D}(\bar{D}^{*||})$, $D_s^{\pm}(D_s^{*\pm||})$. The modified masses lead to an appreciable change in the partial decay widths for $D^*\rightarrow D\pi$ (along with the charge conjugate modes) and $\Psi(3770)\rightarrow D\bar{D}$. The in-medium masses and decay widths thus obtained, accounting for the Dirac sea effects, should modify the yields of the open charm and charmonium mesons arising from the peripheral ultra-relativistic heavy ion collision experiments, where huge magnetic fields can be generated. 
\acknowledgements
Amruta Mishra acknowledges financial support from Department of Science and Technology (DST), Government of India (project no. CRG/2018/002226).

\end{document}